\documentclass{mn2e}
\usepackage{times}
\usepackage{amsfonts}
\usepackage{graphicx}

\title[]{Dynamical Effect of the Turbulence of IGM on the Baryon Fraction Distribution}

\author[Zhu, Feng \& Fang]
{Weishan Zhu$^{1,2}$\thanks{wszhu@pmo.ac.cn},
Long-Long Feng$^{1}$,
Li-Zhi Fang$^{2}$\\
$^{1}$Purple Mountain Observatory, Nanjing, 210008, China\\
$^{2}$Department of Physics, University of Arizona,
Tucson, AZ 85721
}
\begin{document}

\pagerange{\pageref{firstpage}--\pageref{lastpage}} \pubyear{2010}

\maketitle

\label{firstpage}

\begin{abstract}

We investigate the dynamical effect of the turbulence in baryonic
intergalactic medium (IGM) on the baryon fraction distribution. In
the fully developed nonlinear regime, the IGM will evolve into the
state of turbulence, containing strong and curved shocks, vorticity
and complex structures. Turbulence would lead to the density and
velocity fields of the IGM to be different from those of underlying
collisionless dark matter. Consequently, the baryon fraction $f_b$
will deviate from its cosmic mean $f_b^{\rm cosmic}$. We study these
phenomena with simulation samples produced by the weighted
essentially non-oscillatory (WENO) hybrid cosmological
hydrodynamic/N-body code, which is effective of capturing shocks and
complex structures. We find that the distribution of baryon fraction
is highly nonuniform on scales from hundreds kpc to a few of Mpc, and
$f_b$ varies from as low as 1\% to a few times of the cosmic mean.
We further show that the turbulence pressure in the IGM is weakly
scale-dependent and comparable to the gravitational energy density
of halos with mass around $10^{11}$ h$^{-1}$ M$_{\odot}$. The
baryon fraction in halos with mass equal to or smaller than
$10^{11}$ h$^{-1}$ M$_{\odot}$ should be substantially lower than
$f_b^{\rm cosmic}$. Numerical results show that $f_b$ is decreasing
from $0.8f_b^{\rm cosmic}$ at halo mass scales around $10^{12}$ $h^{-1} $
M$_{\odot}$ to $0.3f_b^{\rm cosmic}$ at $10^{11}$ h$^{-1}$
M$_{\odot}$ and shows further decrease when halo mass is less than
$10^{11}$ h$^{-1}$ M$_{\odot}$. The strong mass dependence of $f_b$
is similar to the observed results. Although the simulated $f_{b}$
in halos are higher than the observed value by a factor of 2, the turbulence
of the IGM should be an important dynamical reason leading to the
remarkable missing of baryonic matter in halos with mass $\leq
10^{12}$ h$^{-1}$ M$_{\odot}$.

\end{abstract}


\section[]{Introduction}

In the concordance $\Lambda$CDM universe, the baryonic gas (IGM)
traces the collisionless cold dark matter in the linear regime of
gravitational clustering. However, observation shows that these two
components do not related to each other by a linear mapping on
scales up to a few Mpc. Most attention on the reason of the IGM-cold
dark matter separation on such scales has been drawn to the thermal
property of baryonic gas, mainly radiative heating and cooling, such
as photo-heating, feedbacks from star formation and accretion to
black holes (e.g., Valageas \& Silk 1999; Tozzi \& Norman 2001; Voit
et al. 2002; Zhang \& Pen 2003; Xue \& Wu 2003).

The hydrodynamical origin of the IGM-dark matter separation has been
noticed in the early studies of large scale structure formation
(Shandarin \& Zel'dovich 1989). Successive studies found that in
the weak and moderate nonlinear regime of clustering, the evolution
of collisionless particle can be described by the Zel'dovich
approximation, while the dynamics of baryonic component is sketched by
the adhesion approximation with dissipation of kinetic energy, or by
a random force driving Burgers equation (Gurbatov, Saichev, \&
Shandarin 1989; Berera \& Fang 1994; Vergassola et al. 1994; Jones
1999; Matarrese \& Mohayaee 2002; Pando, Feng, \&
Fang 2004). When the Reynolds number is large, Burgers fluid will be
turbulent, consisting of shocks (L\"assig 2000).  Burgers shocks
occur not only in high, but also in middle and even low density
regions. These shocks could contribute partly to the deviation of the
velocity and density fields of the IGM from that of cold dark matter (Pando et
al 2004; Kim et al 2005).

A new progress shows that in the highly developed nonlinear regime
the velocity field of the IGM is no longer potential, but dominated
by vorticity on scales from  one and a half
hundred kpc to a couple of Mpc (Zhu, Feng \& Fang 2010). Oblique shocks will form in
inhomogeneous baryonic fluid and act as the source of vorticity of
the IGM. These results give supports to the scenario that in the
nonlinear regime the baryonic fluid is in the state of fully developed
turbulence, which can be characterized by the She-Leveque' (SL)
scaling (He et al. 2006) and the log-Poisson hierarchy (Liu \& Fang
2008). Although tiny vorticity can also be generated in dark matter
field during shell crossing, the velocity field of dark matter is
still dominated by potential motion on scales around $1$ Mpc (Pichon
\& Bernardeau 1999; Pueblas \& Scoccimarro 2008). Consequently, the
velocity and density fields of the IGM would depart from those of
underlying collisionless dark matter on scales below a few Mpc. Some
features predicted from the fully developed turbulence have been
found to be consistent with observation, such as the
log-Poisson non-Gaussianity of Ly$\alpha$ transmitted flux of
quasar's absorption spectrum (Lu , Chu, \& Fang 2009; Lu et al.
2010), the scaling relations among the X-ray luminosity,
temperature and SZ effect of clusters (Zhang et al 2006; Yuan et
al. 2009) and the turbulence broadening of HI and HeII Ly$\alpha$
absorption lines (Zheng et al. 2004; Liu et al. 2006).

In this paper, we extend these studies to the baryon fraction
$f_{b}$, which is defined as the mass ratio between the baryonic and
total matter in a system.  In the concordance $\Lambda$CDM universe,
the cosmic mean of baryon fraction is $f_{b}^{\rm
cosmic}=\Omega_{b}/\Omega_{m}=0.17\pm0.01 $, where $\Omega_{b}$ and
$\Omega_{m}$ are the mean mass density parameters of the baryonic
matter and total matter respectively (Dunkley et al. 2009; Komatsu
et al. 2009). The baryon fraction in gravitational bound objects
is found to be lower than the cosmic mean, known as the
missing baryon problem. For galaxy clusters and groups, the
baryon fraction is smaller than the cosmic mean by a factor of 2 - 4
(Ettori 2003; Giodini et al. 2009; Dai et al. 2010), while can be as low as
about $1\%$ of the cosmic mean for dwarf galaxies(McGaugh et al. 2010).
Generally, the baryon fraction
decreases monotonically with decreasing mass of collapsed halos. This
problem has also been seen in the study of  galaxies abundance. The
$\Lambda$CDM model predicts too many low mass dark matter halos in
comparison with estimation from the observed luminosity function
of galaxies. It implies that the star formation rate in low mass
dark matter halos is substantially low. The baryon content residing
in virialized objects with small mass should be much less than that
given by the cosmic mean. Additional physics are needed to keep the
baryonic matter from overcooling (White \& Frenk 1991).

Many kinds of mechanisms have been proposed to prevent most of
the baryon falling into dark matter halos. Feedback from supernovae
and AGN and photo-heating are well investigated. Feedback from
massive stars and SNe were once believed to be able to driving out
the collapsed gaseous baryon and hence suppress the star
formation(Dekel \& Silk 1986; White \& Frenk 1991; Kauffmann et al.
1999). Theoretically, simulations and observations have shown that
this mechanism is unlike to work while taking appropriate feedback efficiency, mass
loss rate(e.g. Mac Low \& Ferrara 1999; Benson et al. 2003), except
for low mass halos.

In contrast, mechanisms in which the missing baryon are inhibited to
fall into the protogalactic halos in the very beginning, i.e., never
fell into, are much preferred (Mo et al. 2005; Anderson \& Bregman
2010). Photo-heating fulfills this category which adopt a
photo-ionizing field to reheat the baryon around progalactic halos
and hence keep them from collapse into potential wells(e.g.
Efstathiou 1992; Gnedin 2000; Benson 2002). The
efficiency of this model, however,  was later found to be largely limited,
only works for halos below $~10^{10}$  M${\odot}$ (Hoeft et al.
2006), because the central self-shielding will delay the photo-evaporation.

In our view, IGM turbulence should be an important reason of the missing baryon
problem. The effects of turbulence on the properties of clusters have been
 extensively studied by numerical simulations and theoretical works(e.g. Norman \&
Bryan 1999; Dolag et al. 2005; Vazza et al. 2009, 2010; Burns,
Skillman\& O'Shea 2010;  Ruszkowski \& Oh 2010). Yet, we will focus
on the dynamical and statistical effect of turbulence on the spatial
distribution of baryon fraction. That is, the turbulence of IGM is
treated as a basic environment factor of gravitational clustering.
The baryon missing in virialized objects should be a result of the
formation and evolution of inhomogeneity of  $f_b$ spatial
distribution. One can then explain the baryon missing by considering
the inhibition of gravitational collapse by turbulence pressure
(Chandrasekhar 1951a, 1951b).

This paper is organized as follows. \S 2 addresses the theoretic
background of turbulence in the IGM and its effect
on the gravitational collapsing of baryonic matter. \S 3
gives the simulation method of producing samples.
In \S 4, we study the properties of the non-uniform distribution of
the baryon fraction, and its relation with turbulence. \S 5 presents
the baryon fraction in collapsed halos. We discuss our results and compare 
them with previous numerical studies in \S 6. Finally, the conclusions are given in \S 7. 

\section[]{Theoretical background}

\subsection{Turbulent IGM}

Hierarchical structure formation process will introduce shocks(Ryu et al.
2003; Pfrommer et al. 2006; Vazza et al. 2009). The  shock wakes and
shear flows, arose from gas accreting into pancakes, filaments and halos, protogalactic
and collapsed objects moving in complex structures  and gaseous
structure colliding and merging, will interplay with each other
and drive instability, result in the onset of turbulence.
Fully developed turbulence consists of eddies on various scales and will
cascade the kinetic energy of largest eddies down to smaller
one (Lin 1966; Shu 1992).

More specifically, the evolution of baryonic fluid in the moderate
nonlinear regime can be characterized by random force driving
Burgers equation (Gurbatov et al. 1989; Berera \& Fang, 1994; Jones,
1999; Matarrese \& Mohayaee, 2002). When the effective Reynolds
number is large, burgers fluid will be turbulent, consisting of
shocks and complex structures on various scales(Polyakov, 1995;
L\"{a}ssig 2000; Boldyrev et al. 2004). As dark matter is not
influenced by Burgers turbulence, the IGM velocity field will
dynamically decouple from the dark matter field on scales larger
than the Jeans length once Burgers turbulence is developed. This
will lead to the deviation of $f_b$ from $f_{b}^{\rm cosmic}$ in low
and high density areas. 

The velocity field of baryonic fluid keeps irrotational in the moderate nonlinear
regime.  Latter on, the velocity field of IGM will no longer be potential dominated,
because the vorticity can be
effectively generated by oblique shocks due to baroclinic instability(Zhu et al 2010).
When shocks propagate in inhomogeneous medium, they will generally evolve into oblique or
curved shock waves (e.g. Landau \& Lifshitz 1987), acting as
the source of vorticity (e.g. Picone et al. 1984; Emanuel 2000).
Once triggered, the vorticity can be self-amplified by
the nonlinearity of hydrodynamics (see \S 2.3). Finally the IGM evolves
into a fully developed turbulent state which satisfies the
She-Leveque scaling and log-Poisson hierarchy (He et al. 2006; Liu \&
Fang 2008; Lu et al 2009, 2010). More details about the
development of turbulence in the IGM  along cosmit evolution
can be found in (Zhu et al. 2010; Fang \& Zhu 2011).

\subsection{Turbulence pressure and vorticity of IGM}

The Jeans length of  self-gravitational clustering in a statistical
uniform gas with density $\rho$ and speed of sound $c_s$  with or
without turbulent gas motion has been studied by Chandrasekhar (1951a, 1951b).
In the absence of turbulence, the Jeans length is $\lambda_J=c_s\sqrt{\pi/G\rho}$,
i.e. gravitational clustering can occur only if the perturbation
scale is larger than $\lambda_J$. The effect of turbulent motions on
the clustering is to replace $c_s$ by an effective sound speed
\begin{equation}
c^2_{\rm s,eff}=c_s^2+\frac{1}{3} \langle v^2\rangle
\end{equation}
where $\langle v^2\rangle$ measures the velocity fluctuations of
turbulence. The Jeans length is increased by the random
velocity field of turbulence, which plays the similar role as the randomly
thermal motion. The random motion of the turbulence
provides an extra pressure, $p_{\rm tub}=\langle \rho v^2\rangle$,
called turbulence pressure. The turbulence pressure, joining the
thermal pressure, will slow down and even halt
the IGM falling into a gravitational well.

Considering the gravitational collapsing on scale $R$ will not be
affected by the velocity dispersion on scales that larger than $R$, the
velocity fluctuations with wave-numbers $k< 2\pi /R$ will not resist
gravitational collapsing on scales larger than $R$. Consequently, the effect of
turbulence pressure on gravitational collapsing on scale $R$ could
be estimated by (Bonazzola et al 1987)
\begin{equation}
p_{\rm tur}=\int_{\max(k_{\rm R}, k_{\rm min})}^{k_{\rm max}}E(k)
dk,
\end{equation}
where $E(k)$ is the power spectrum of kinetic energy density
$(1/2)\rho({\bf r})v^2({\bf r})$ of the turbulence. The upper limit
of the integral eq.(2)  ${k_{\rm max}=2\pi/l_{\rm diss}}$
corresponds to the minimal scale $l_{\rm diss}$ below which the
turbulence is dissipated. The lower limit of the integral eq.(2) is
 the maximum of $k_{\rm R}$ and $k_{\rm min}$, where
$k_{\rm R}=2\pi/R$ and $k_{\rm min}$ matches the upper scale
of turbulence. Obviously, the
turbulent pressure $p_{\rm tur}$ is dynamical, not thermal. It can
be comparable to the thermal pressure of the IGM, especially in regions
the temperature of IGM is not very high.

When we estimate the turbulence pressure with eq.(1) or (2), we
should separate the velocity field ${\bf v}({\bf r},t)$ into two
components: one is the random motion of turbulence and the other is
the bulk velocity. The latter mainly depends on the
gravitational potential, and does not contribute to the turbulence
pressure. In some algorithms, the bulk velocity is identified as the
mean velocity within a box, while the fluctuation with respect to
the mean velocity is supposed to be the turbulent motion.
As the box size is selected by hand, these algorithms might contain
non-ignorable system bias. A better way to pick up the random motion of
turbulence bases on the vorticity of  velocity field.

The vorticity field $\vec{\omega}({\bf r},t)$ of the velocity field
${\bf v}$ of the IGM is defined by
$\omega_i=(1/2)(\partial_iv_j-\partial_j v_i)$, i.e.,
$\vec{\omega}=\nabla \times {\bf v}$, where $i=1,2,3$. The dynamical
equation of the vorticity $\vec{\omega}$ is (Zhu et al. 2010)
\begin{eqnarray}
\frac{D \vec {\omega}}{Dt} & \equiv & \partial_t {\vec \omega}
+\frac{1}{a}{\bf v}\cdot{\nabla}\vec{\omega} \\
\nonumber
& = &\frac{1}{a}({\bf S}\cdot {\vec\omega}-d {\vec\omega}
+\frac{1}{\rho^2}\nabla \rho \times\nabla p-\dot{a}\vec{\omega}),
\end{eqnarray}
where $p$ is the pressure of the IGM, $d=\partial_iv_i$ is the
divergence of the velocity field and $a(t)$ is the cosmic factor.
Tensor ${\bf S}$, defined as
$S_{ij}=(1/2)(\partial_iv_j+\partial_jv_i)$, is called strain rate,
and $[{\bf S}\cdot {\vec\omega}]_i=S_{ij}\omega_j$. A
remarkable property of eq.(3) is the absence of the gravity term
. This point is expected, as gravity is curl-free in
nature. In other words, vorticity is fully given by the nonlinearity of
hydrodynamics and hence is effective to measure the
turbulence.

An important property of the vorticity is that for a fully developed
turbulence the power spectra of the vorticity field $P_{\omega}(k)$
and the velocity field $P_{v}(k)$ should satisfy
$P_{\omega}(k)=k^2P_{v}({\bf r})$ (Batchelor 1959, 2000). This
relation can be used to determine the wavenumber $k_{\rm min}$ , i.e.  the
upper scale of fully developed turbulence. We use this criterion to
ascertain $k_{\rm min}$ to calculate the integral eq.(2).

\subsection{Dynamical effect of turbulence on IGM gravitational clustering}

The dynamical effect of turbulence on the gravitational clustering
of the IGM can be seen with the time evolution of the irrotational
component of velocity field, i.e. the divergence of velocity field
$d=\partial_iv_i$, which follows (Zhu et al 2010)
\begin{eqnarray}
-\frac{D d}{D t} &\equiv & -\partial_td  - \frac{1}{a}{\bf v}\cdot{\bf \nabla}d = \\
\nonumber
 & = & \frac{1}{a}\left [S_{ij}S_{ij}-\frac{1}{2}\omega^2
+\frac{1}{\rho}\nabla^2 p
-\frac{1}{\rho^2}(\nabla\rho)\cdot(\nabla p)   \right . \\
\nonumber &  & +\left . \frac{4\pi G}{a}(\rho_{\rm
tot}-\rho_0)+\dot{a}d \right ],
\end{eqnarray}
where $\rho_{\rm tot}$ is the total mass density including both cold dark
and baryonic matter with $\rho_{0}$ being its mean value. As negative divergence
means the increase of density,  positive  $4\pi
G(\rho_{tot}-\rho_0)/a$ will lead to clustering, while negative one
result in anti-clustering. The right hand side of eq.(4) can be used to
compare the effects of hydrodynamical terms on clustering with that of
gravity.

We first identify the physical meaning of the hydrodynamical terms
on the right hand sight of eq.(4) by considering incompressible
fluid, i.e. assuming $\rho_{\rm tot}$  to be a constant $\rho_{\rm tot}=\rho_0$. In
this case, the term of gravity $(4\pi/a)(\rho_{\rm tot}-\rho_0)=0$,
and $\nabla \rho=0$. The continuity equation yields $d=0$, and
Eq.(4) gives
\begin{equation}
\nabla^2 p=-\rho\left (S_{ij}S_{ij}-\frac{1}{2}\omega^2 \right ).
\end{equation}
This is a typical Poisson equation for the scalar field $p$.
Analogous to the field equations in
electrostatics, the term $\rho[S_{ij}S_{ij}-{1/2}\omega^2]$
on the right hand side of eq.(5) plays the role of the ''charge" of a
pressure field.  Positive ''charge" produces an attraction force,
while  negative ''charge" yields a repulsive force. Back to
eq.(4), $\rho [S_{ij}S_{ij}-{1/2}\omega^2]$ also plays the
role of nonthermal pressure of turbulence. In regions with negative
''charge", i.e. $[S_{ij}S_{ij}-(1/2)\omega^2 ]< 0$, the turbulent
fluid will prevent the gravitational clustering.

The sign of the ''charge" is actually determined by the levels that
turbulence has developed. From the definition of vorticity and
strain rate, we have
\begin{equation}
\frac{1}{2}\omega^2- S_{ij}S_{ij}=\frac {1}{2}
[(\partial_iv_j)(\partial_iv_j)-3(\partial_j v_i)(\partial_iv_j)].
\end{equation}
For a Gaussian velocity field, $\langle 3(\partial_j
v_i)(\partial_iv_j)\rangle=\langle
(\partial_iv_j)(\partial_iv_j)\rangle$,  the net
effect of velocity fluctuations on the IGM collapsing is null in
average. However, for a non-Gaussian velocity field, it can be
either positive or negative, depending on the property of
the velocity field. For a homogeneous and isotropic turbulence
, $\langle (\partial_iv_j)(\partial_jv_i)\rangle=0$ (Batchelor
1959), the sign of the ''charge", $\rho
[S_{ij}S_{ij}-{1/2}\omega^2]$, is always negative.  A fully
developed turbulent flow will generally prevent the IGM from
gravitational collapsing.

The term $\nabla^2 p/\rho$ of eq.(4) relates to the hydrostatic pressure
while does not consider fluid compressibility, which has been presented in eq.(5).
It  is mostly negative in overdense collapsing regions, and will resist
upon gravitational collapse. The term $-(\nabla\rho)\cdot(\nabla
p)/\rho^2$ on the right hand side of eq.(4) result from the compressibility
of the IGM.  Its value would be negative when the density-pressure
relation is a power law $p\propto \rho^{\gamma}$
and $\gamma>0$, and then also plays the role of resisting
gravitational collapsing.

The physical meaning of the term $(\nabla\rho)\cdot(\nabla
p)/\rho^2$ can be shown by the ratio between
$(\nabla\rho)\cdot(\nabla p)/\rho^2$ and the gravity term $4\pi
G(\rho_{\rm tot}-\rho_0)$. The ratio is roughly equal to $\sim
(t_{\rm infall}/t_{\rm sound})^2$, where $t_{\rm infall}\sim
(G\rho)^{1/2}$ is the free falling time scale, $t_{\rm
sound}\sim R/c_s$, $c_s\sim (\nabla p/\nabla \rho)^{1/2}$, is the
dynamical time scale on a spatial scale $R$ collapsing.
When $(t_{\rm infall}/t_{\rm sound})^2$ is larger than $1.0$, the free
falling time scale is larger than the dynamical time scale, and in result,
the collapsing on scale $R$ is significantly prevented.

\section[]{Method}

As mentioned above, the turbulent IGM contains curved shocks,
vortices, and other discontinuities. To study the effect of turbulent IGM with
simulation, the algorithm should be qualify for
capturing these complex structures. We take advantage of the
development of the Weighted Essentially Non-oscillatory (WENO)
method (Shu 1999). The WENO schemes have been widely used in various
fields, such as high Reynolds number compressible flows (Zhang et
al. 2003), high Mach number jets (Carrillo et al. 2003),
magneto-hydrodynamics (Jiang \& Wu, 1999) and hypersonic boundary layer
(Rehman et al 2009). The shock capturing algorithm with WENO scheme
has past many test, including shock-boundary
layer interaction (Lagha et al. 2009), shocks in high-speed flows
(Martin, Piomelli \& Candler 2000), shock vortex interaction (Grasso
\& Pirzzoli 2000a, 2000b) and shock-turbulence interaction (Pirozzoll
2002). An updated review of the WENO method is given in Shu 2009.

In the context of cosmological hydrodynamical simulation, the WIGEON
code is based on Eulerian description of hydrodynamics with 5th
order WENO finite difference scheme and particle mesh (PM) method
for dark matter particles (Feng, Shu, \& Zhang 2004). The WIGEON code can
reproduce commonly accepted results such as the relation between
IGM temperature and density and the component of WHIM (He, Feng, \& Fang 2004). In
the same time, it also reveals a series of turbulent behavior of
the IGM (He at al 2006; Liu \& Fang 2008; Lu et al. 2009, 2010).
These features of this code fit well with our goal.

The cosmological parameters are taken to be  $(\Omega_{m},
\Omega_{\Lambda}, h, \sigma_{8}, \Omega_{b}, n_{s}, z_{re}) =$
(0.274, 0.726,  0.705, 0.812, 0.0456, 0.96, 11.0) (Komatsu et al.,
2009). The simulation run in a periodic cubic box of size 25
$h^{-1}$ Mpc since redshift $z=99$ with a  $512^{3}$ grid and an
equal number of dark matter particles, giving a mass resolution of
$1.04 \times 10^{7} M_{\odot}$. A uniform UV background of ionizing
photons is added at $z_{\rm re}$ to mimic the reionization.
Radiative cooling and heating are followed as Theuns et al.
(1998) with the primordial composition $X=0.76, Y=0.24$.  Star
formation and its feedback are not added in our simulation, because the resolution
we used makes it hard to implant these processes appropriately.
Snapshots are outputted at redshifts
$z=11.0,6.0,4.0,3.0,2.0,1.0,0.5,0.0$. To study the convergence of
numerical results, we also run a simulation with $256^{3}$ grid and
an equal number of dark matter particles.

The velocity field of  simulation samples has been used
to show the development of vorticity and turbulence in Zhu et al.(2010).
To locate shocks post simulation we use a algorithm also based on WENO
kernel, combining with conditions from Ryu et al.(2003) (see Appendix). To
construct the density field of dark matter on grids, we assign the
mass of dark matter particles onto grids using the
Triangle-Shaped-Cloud (TSC) method. In order to minimize the system bias
of assignment in calculating the baryon fraction on grids, we
smooth over the density fields of baryonic and dark matter on grids
separately using the same smooth window with radius of one grid.

To identify halos of dark matter and their radius $r_{200}$, we use
the same process as Crain et al. (2007). Halos
are identified by two methods: a.) friends-of-friends (FOF) method
with a linking length parameter $0.2$; b.) HOP method
(Eisenstein \& Hut 1998) with default parameters for group searching
and  $\delta_{\rm outer}=80$, $\delta_{\rm saddle}=200$ and
$\delta_{\rm peak}=240$ for group merging. We study only halos
consisting of no less than 2000 dark matter particles at $z=0$. For each halo,
we can then find the center and radius of a sphere, in which the
mean mass density is equal to $200 \rho_{\rm crit}(z)$, where
$\rho_{\rm crit}(z)=3H^2(z)/8\pi G$ is the critical density at
redshift $z$.

\section[]{Spatial distribution of baryon fraction}

\subsection{A Slice}

\begin{figure*}
\begin{center}
\includegraphics[width=8.0cm]{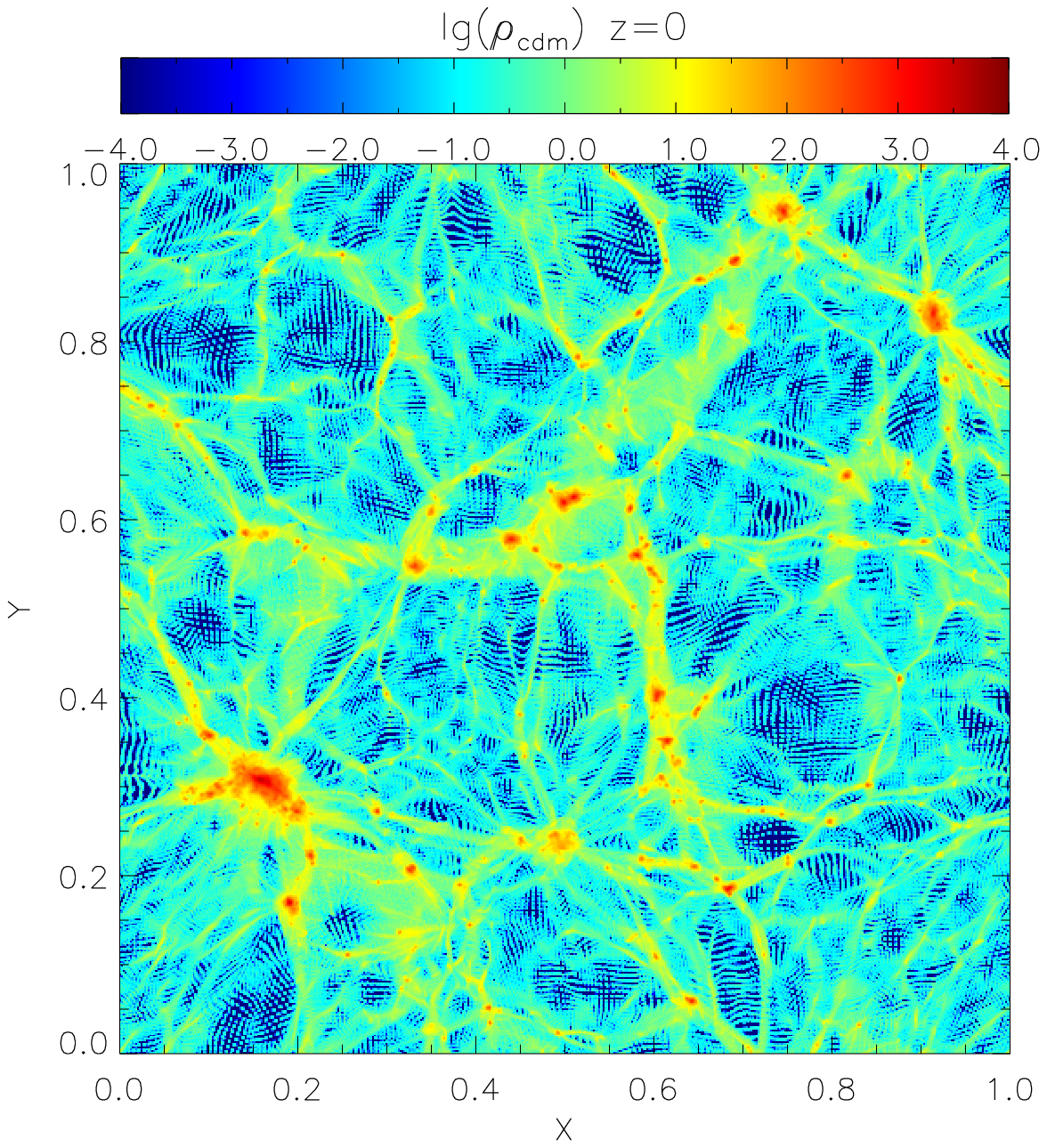}
\includegraphics[width=8.0cm]{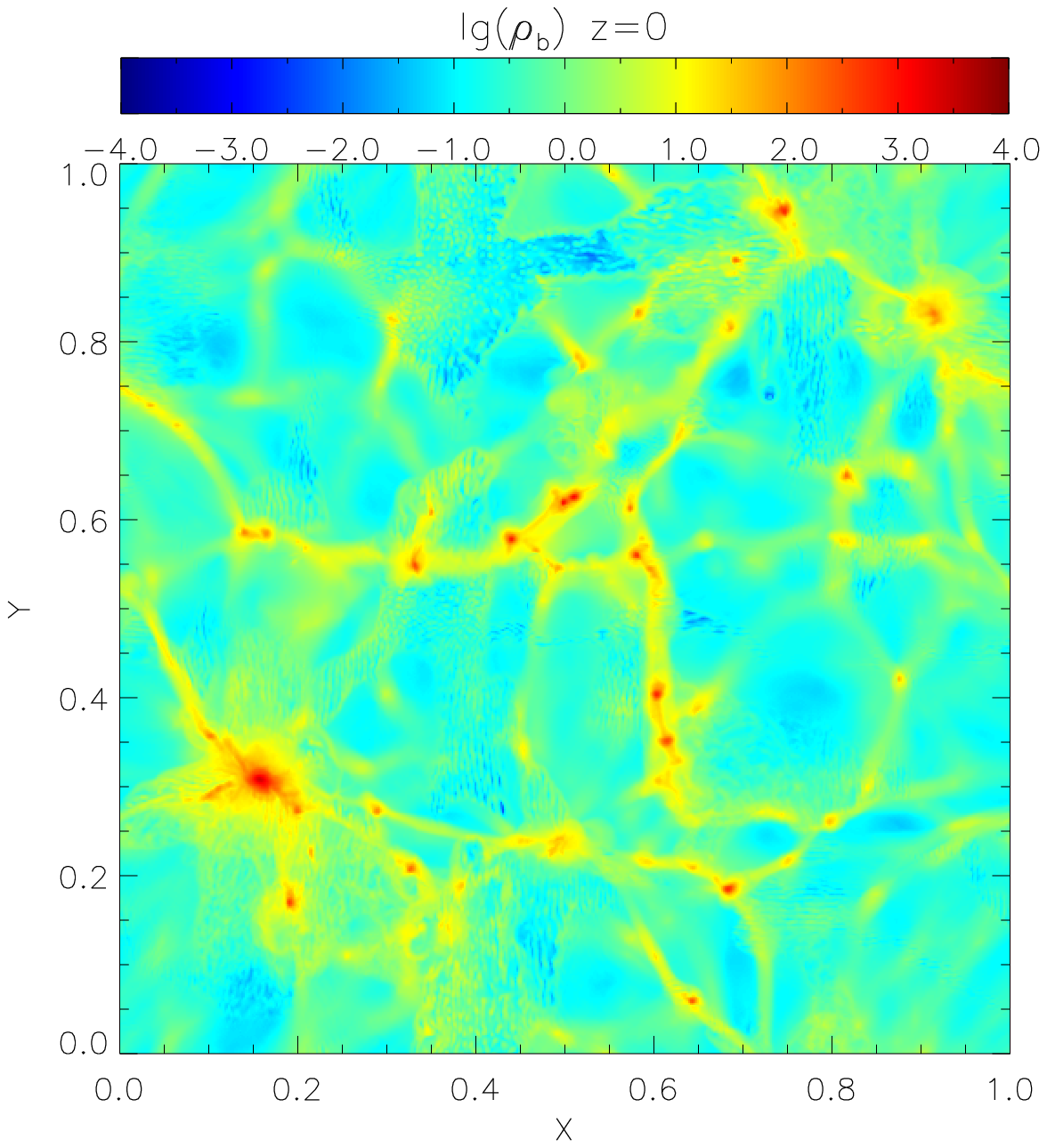}
\end{center}
\begin{center}
\end{center}
\vspace{-7mm}
\caption{Distributions of dark matter density (left); baryon density (right)
in a slice of 25x25x0.2 $h^{-3}$ Mpc$^3$ at redshift $z=0$.}
\end{figure*}

Figure 1 presents an example of the spatial distributions of baryonic and dark matter
density in a slice of 25x25x0.2 $h^{-3}$ Mpc$^3$
at redshift $z=0$. The density fields of dark matter and baryonic fluid of Figure
1 display the typical sheets-filaments-knot structures on the cosmic scales.

The spatial distributions of shock, temperature, vorticity and
normalized baryon fraction $F_b\equiv f_b/f_b^{\rm cosmic}$ on the
same slice are shown in Figure 2. The shocks do not always follow the
filament and sheet structures of the  matter density
fields. Shocks can also be formed in low density areas, as demonstrated
in other simulations (e.g. Ryu et al. 2003; Kang
et al. 2007).

The vorticity field in Figure 2 is described by the dimensionless
quantity $\omega t$, where $t$ is the cosmic time.
$\omega t$ accounts the number of rotation of the vorticity $\omega$ in
the age of universe. The distribution of vorticity also doesn't
always follow the filament and sheet structures of the matter
density fields, but has cloud-like structures (Zhu et al 2010).

Although curved shocks are the sources of vorticity, the spatial
distribution of vorticity field does not show the same configuration
as the shocks. It is because the spatial transfer of vorticity is
given by the velocity field [eq.(3)], while the case with
shock front is different. Once the vorticity appears,
it will depart from their source-curved shocks
and spreads over the space along with the velocity
field (Batchelor 2000).

In the plot of vorticity distribution, we add FOF identified
halos, which are marked as solid circles. The radius of circles are
enlarged to 5 times of the real halos radius. A remarkable
feature is that all the halos are located in the area of $\lg \omega
t \geq 1$. It means that all the gravitational bounded halos are
formed in the environment of turbulent IGM.  As discussed
in \S 2.2, the gravitational clustering should be affected by the
turbulence pressure.

The spatial distribution of the normalized baryon fraction $F_b$ is
highly non-uniform. The value of $F_b$ spreads over three orders
of magnitude from $\lg F_b\sim -2$ to $\sim 1$, indicating that the
separation of baryonic matter from dark matter is significant developed during
the nonlinear evolution. The deviation of baryon fraction from the
cosmic mean is not only occurred in or around massive halos, but
also in the low and moderate density areas. The spatial distribution of baryon
fraction does not display the sheets-filaments-knot structures as
the density fields of dark and baryonic matter do. Nevertheless,
the area of high $\rho_{\rm cdm}$ are surrounded by high
baryon fraction area (dark blue),  indicating that the processes of
separating baryonic matter from dark matter would be potent
around high density areas. The sizes of some region in which $\lg
F_b\leq -0.5$, corresponding to $F_{b} ~ \leq 0.3$, is much larger
than the resolution, demonstrating that the existence of regions with
$F_b \leq 0.3$ is robust.

\begin{figure*}
\begin{center}
\hspace{-0.2cm}
\includegraphics[width=5.5cm]{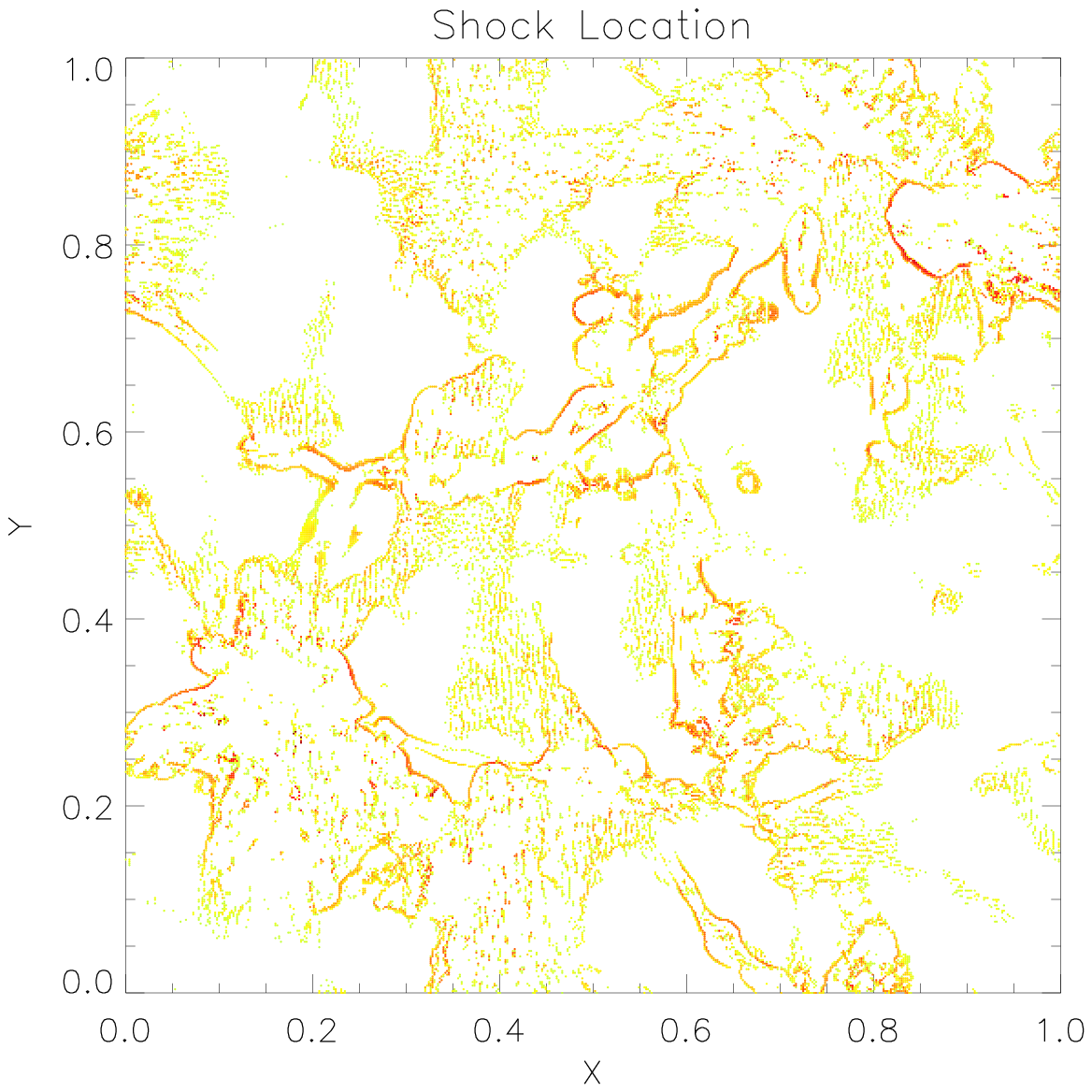}
\hspace{2.5cm}
\includegraphics[width=5.5cm]{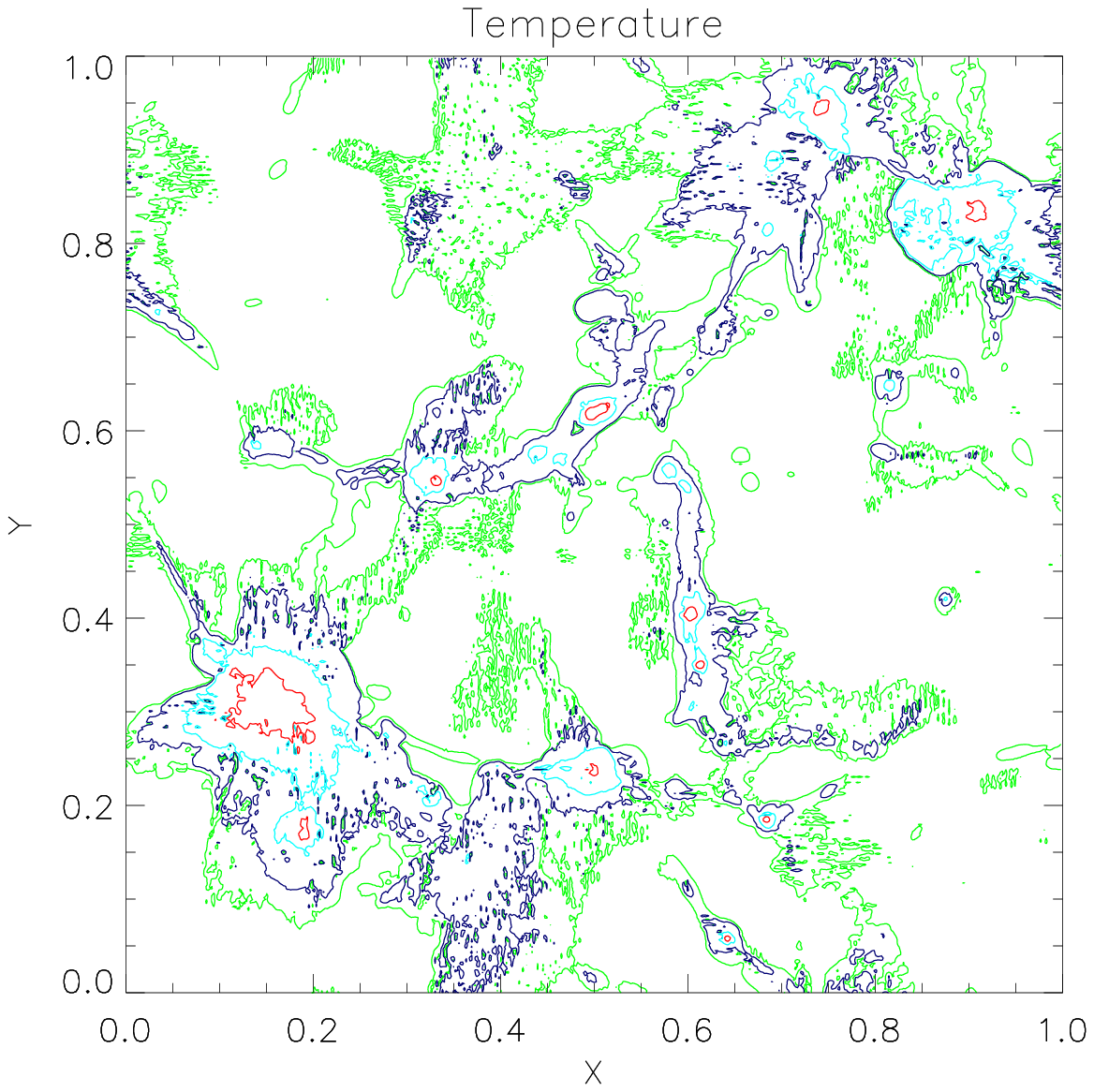}
\end{center}
\begin{center}
\includegraphics[width=8.0cm]{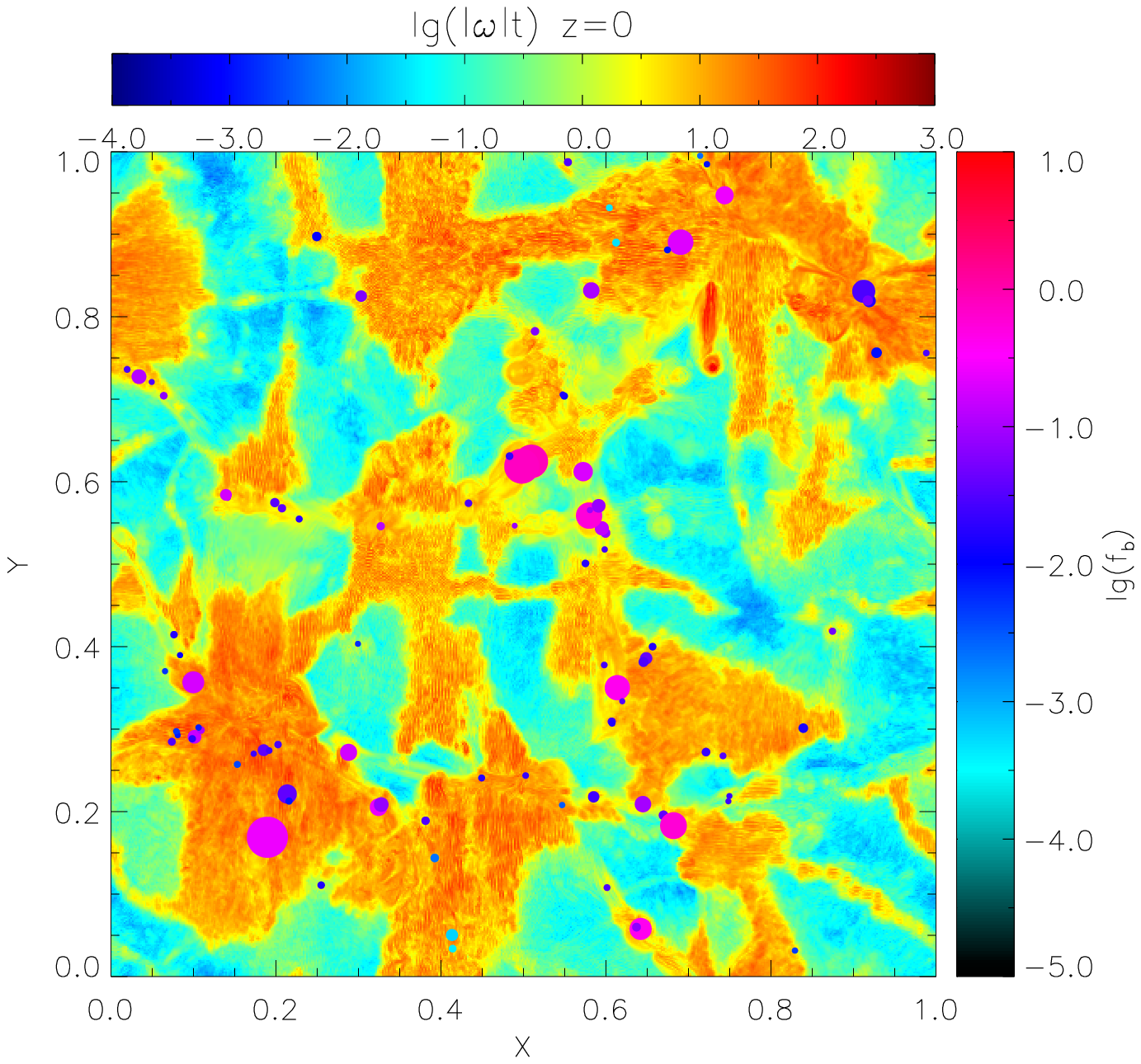}
\hspace{-0.5cm}
\includegraphics[width=8.0cm]{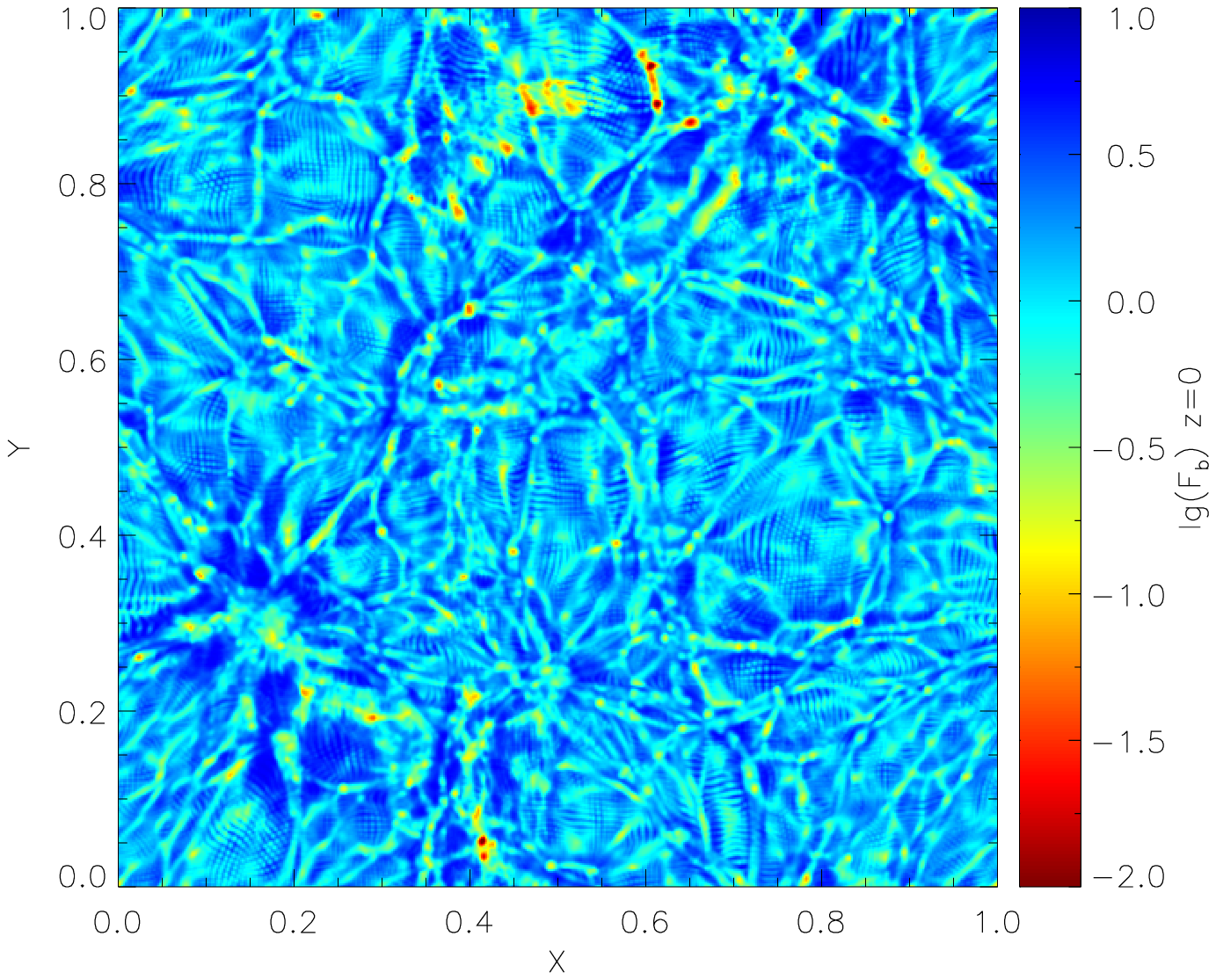}
\end{center}
\vspace{-2mm} \caption{The top left shows shocks with Mach number $
M_a \sim 1-10$ (green), $ \sim 10-100$ (yellow) and $>100$ (red).
The top right is the temperature contours with green, navy, cyan
and red representing 10$^4$, 10$^5$, 10$^6$, 10$^7$ K,
respectively. The bottom left is the vorticity in the slice, in
which the solid circles are halos identified by FOF method. The
radius of each solid circle is equal to 5 times of the halo radius.
The bottom right is the normalized baron fraction
$F_b=f_{b}/f_{b}^{\rm cosmic}$.}
\end{figure*}

\subsection{Comparing the effects of turbulence and gravity on clustering}

As mentioned in \S 2.3, the term $(4\pi G/a)(\rho_{\rm tot}-\rho_0)$
on the right hand side of eq.(4) measures the gravity effect leading to
clustering ($\rho_{\rm tot}>\rho_0$), or anti-clustering ($\rho_{\rm
tot}<\rho_0$). The first three terms on the right hand side of
eq.(4) describe the gas dynamical and thermal effects, namely,
the vorticity and strain rate $-\omega^2/2+
S_{ij}S_{ij}$, the ordinary pressure $\nabla^2 p/\rho$ and
multiphase-related term  $-(\nabla\rho)\cdot(\nabla p)/\rho^2$. We now
use simulation sample to study the general properties of these
terms.
\begin{figure*}
\begin{center}
\includegraphics[width=6cm]{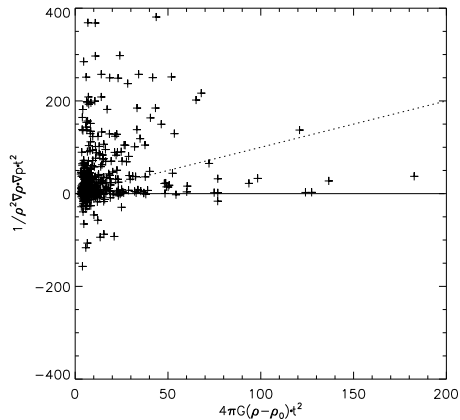}
\end{center}
\vspace{-2mm} \caption{$(\nabla\rho)\cdot(\nabla p)/\rho^2$ vs.
$4\pi G(\rho_{tot}-\rho_0)$ of randomly selected cells at $z=0$. The
dotted line shows  $(\nabla\rho)\cdot(\nabla p)/\rho^2= 4\pi
G(\rho_{tot}-\rho_0).$}
\end{figure*}

Figure 3 presents a comparison between $(\nabla\rho)\cdot(\nabla
p)/\rho^2$ and $4\pi G(\rho_{\rm tot}-\rho_0)$ in cells randomly
selected from the simulation samples at $z=0$. We chose only the cells
of $4\pi G(\rho_{\rm tot}-\rho_0)>0$, which drives collapsing. The
variables used in Fig. 3, $(1/\rho^2)(\nabla\rho)\cdot(\nabla p) t^2$ and $4\pi G(\rho_{\rm
tot}-\rho_0) t^2$, are dimensionless, where $t$ is the age of the universe.

We can see from Figure 3, when $4\pi G(\rho_{\rm tot}-\rho_0)<50$,
more than a half of the data points having
$(\nabla\rho)\cdot(\nabla p)/\rho^2 > 4\pi G(\rho_{\rm
tot}-\rho_0)$ and few points have $(\nabla\rho)\cdot(\nabla
p)/\rho^2<0$, which corresponds to the so-called inverse
density-pressure(temperature) relation. Therefore, the term
$(\nabla\rho)\cdot(\nabla p)/\rho^2$ enhanced by shocks and complex
structures can effectively resist gravitational collapsing till
$4\pi G(\rho_{\rm tot}-\rho_0) \sim 50$. For deep gravity wells with
$4\pi G(\rho_{\rm tot}-\rho_0)>50$, most data points show that
gravity generally is larger than $(\nabla\rho)\cdot(\nabla
p)/\rho^2$. Namely, only in deep gravitational wells the motion of
baryonic matter will be dominated by the gravity.
\begin{figure*}
\begin{center}
\includegraphics[width=6.0cm]{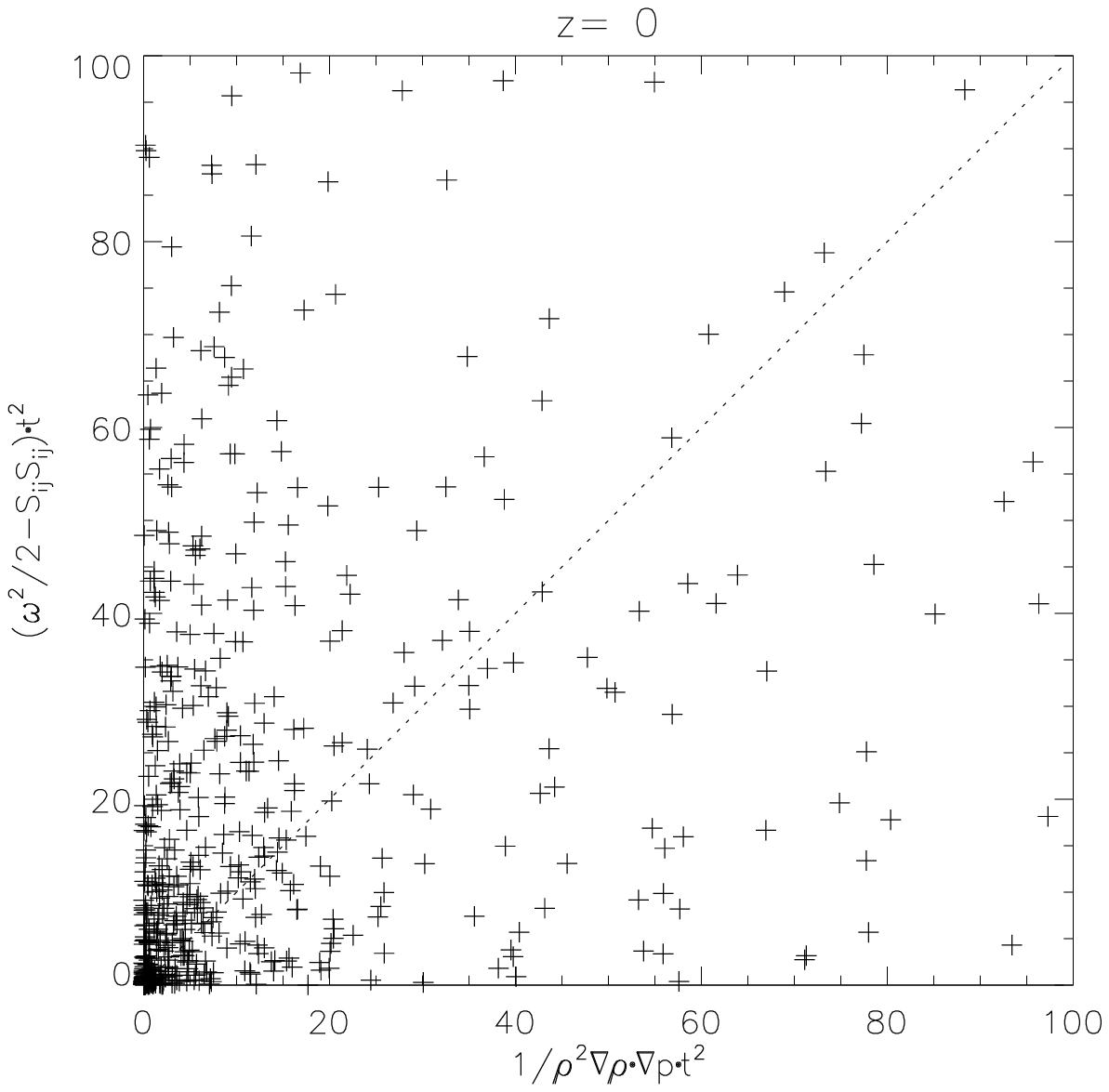}
\hspace{5mm}
\includegraphics[width=6.0cm]{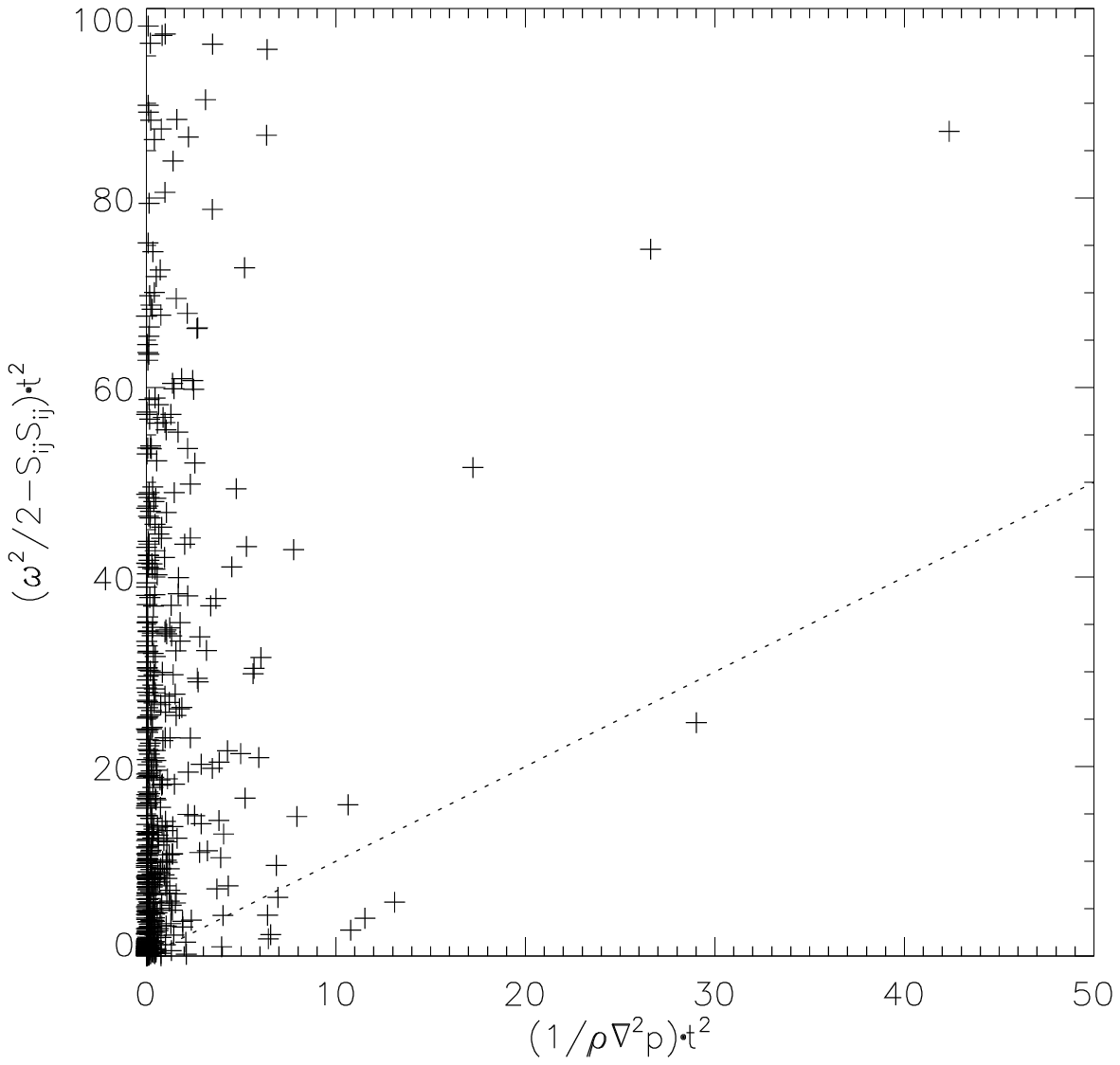}
\end{center}
\vspace{-2mm} \caption{Left panel: $(1/2)\omega^2-S_{ij}S_{ij}$ vs.
$(\nabla\rho)\cdot(\nabla p)/\rho^2$ in randomly selected cells at
redshift $z=0$. Right panel: $(1/2)\omega^2-S_{ij}S_{ij}$ vs.
$(1/\rho)\nabla^2 p$ of the same cells. }
\end{figure*}

Figure 4 compares $\omega^2/2- S_{ij}S_{ij}$ vs.
$(\nabla\rho)\cdot(\nabla p)/\rho^2$ (left panel), and
$(1/2)\omega^2- S_{ij}S_{ij}$ vs.  $(1/\rho)\nabla^2 p$ (right panel)
at randomly selected cells.  We use the dimensionless variables by
multiplying with $t^2$.
The left panel of Figure 4 shows that the magnitude of $\omega^2/2-
S_{ij}S_{ij}$ is comparable with $(\nabla\rho)\cdot(\nabla
p)/\rho^2$.  While in the right panel, it exhibits that term
$(1/2)\omega^2-S_{ij}S_{ij}$ is stronger than the term
$(1/\rho)\nabla^2 p$. On average, the term $(1/\rho)\nabla^2 p$
 is smaller than $(\nabla\rho)\cdot(\nabla p)/\rho^2$.

In a word, in terms of prevention of gravitational collapsing, the effect of
vorticity and strain rate is comparable with
$(\nabla\rho)\cdot(\nabla p)/\rho^2$ and dominates over
$(1/\rho)\nabla^2 p$. Both of the latter two terms are thermal
pressure $p$ related, one can regard vorticity and strain
rate as an effective thermal pressure. Figure 2 shows that in most
regions IGM temperature are equal to or less than 10$^6$ K. Giving
the comparison in the last paragraph, the effective thermal pressure
of turbulence would be as strong as the thermal pressure with
temperature $\sim 10^{5-6}$ K. The turbulence pressure level given
by  simplified estimation is consistent with the more delicate
one using the energy density of turbulence in Zhu et al. (2010).

\subsection{Correlations between baryon fraction and turbulence}

As vorticity is an important dynamical factor to cause a low
baryon fraction, one can expect a correlation between the baryon
fraction and the vorticity. Figure 5 gives the probability density function(PDF)
of $\lg [(\omega t)^2/2]$ in various ranges of the
baryon fraction $F_b$. Figure 5 indeed shows that for cells have
small $F_b$, the PDF at the side of $\lg [(\omega t)^2/2]
>2$ are higher. The cells with more serious baryon missing have larger
vorticities in average.
\begin{figure*}
\centering
\includegraphics[width=6.8cm]{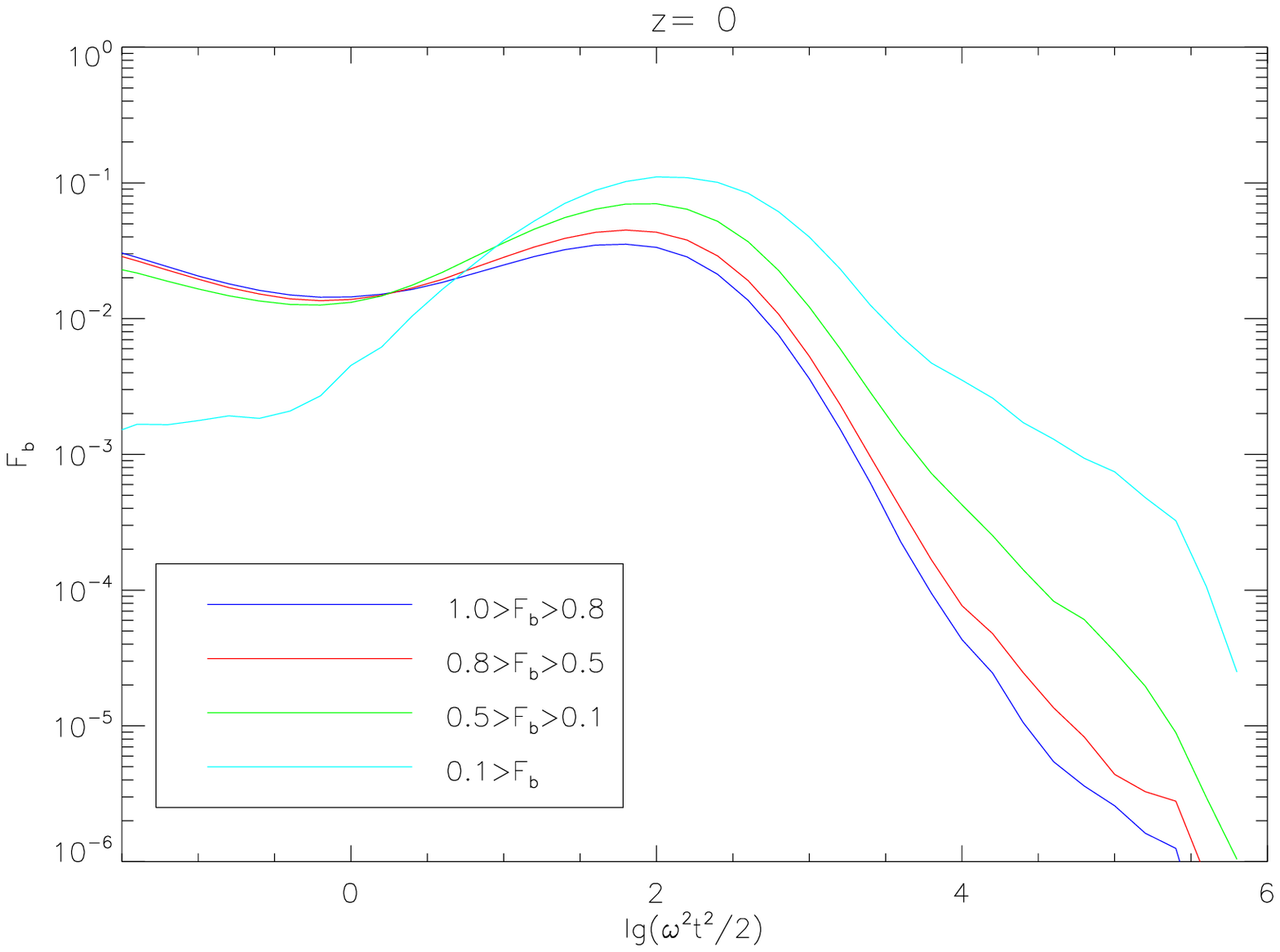}
\hspace{2mm}
\includegraphics[width=6.8cm]{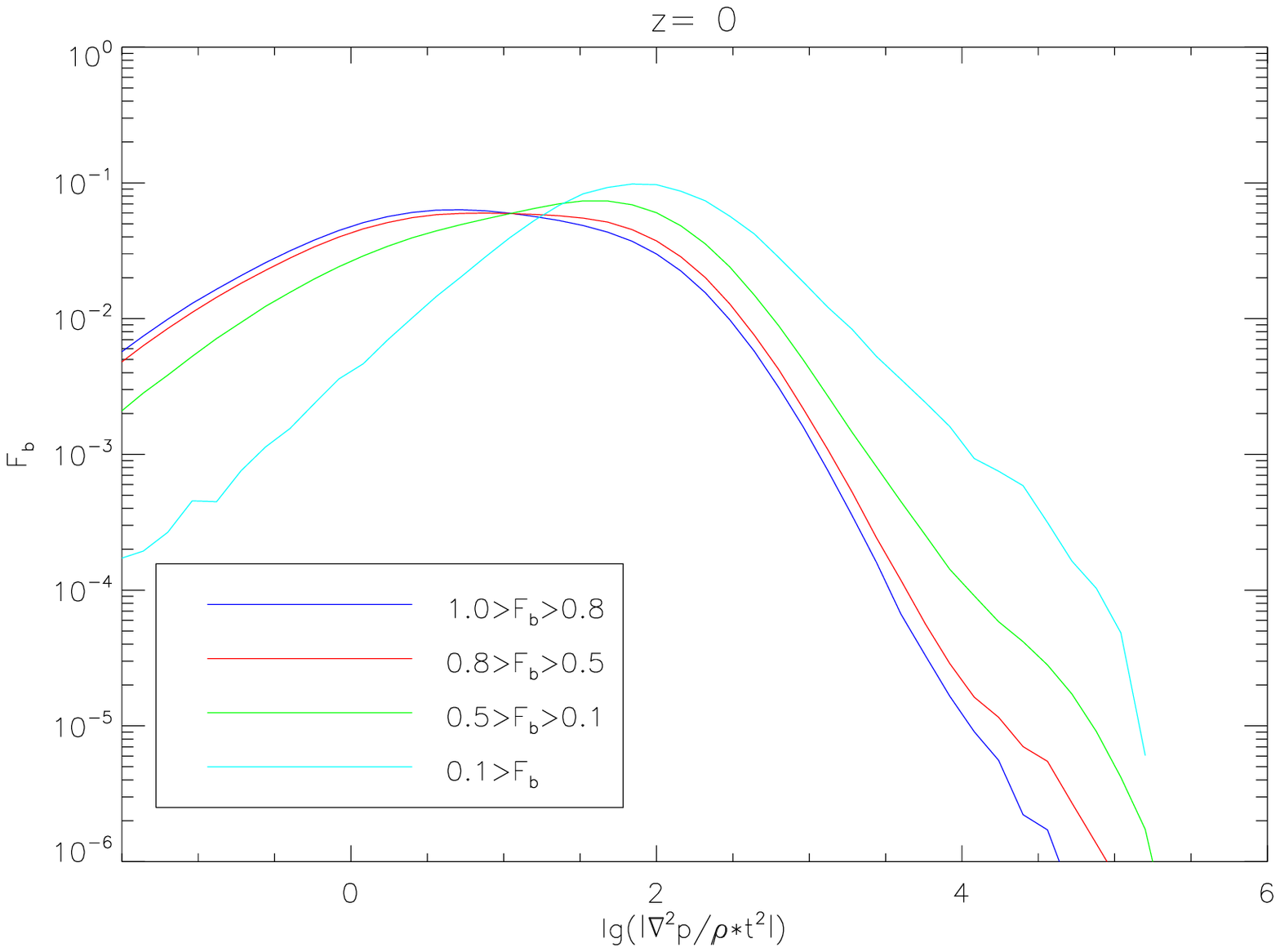}
\caption{The PDF of $\lg|\omega^{2} t^{2}|$ (left) and $\lg|\nabla^2 p/\rho t^{2}|$
(right) for four ranges of $F_b$: a.) $0.8-1.0$ (blue); b.)
$0.5-0.8$ (red); c.) $0.1-0.5$ (green), and d.) $<0.1$ (cyan) for
sample at redshift $z=0$.}
\end{figure*}

The similar PDF of $\lg|(1/\rho)(\nabla^2 p) t^2|$ is also shown in
Figure 5. The cells with more significant baron missing generally
corresponds to large $|(1/\rho)(\nabla^2 p) t^{2}|$ as well.
Although the two panels of Figure 5 look similar, the values of
$(\omega t)^2/2 t^2$ and $ |(1/\rho)(\nabla^2 p) t^2|$ actually
cover different magnitude range. The former can be as large as $10^6$, while
the later is about $10^5$. This is consistent with the right panel
of Figure 4. The vorticity term $(\omega t)^2/2$ is larger than the
term $(1/\rho)(\nabla^2 p) t^{2}$.

\begin{figure*} \centering
\includegraphics[width=6.5cm]{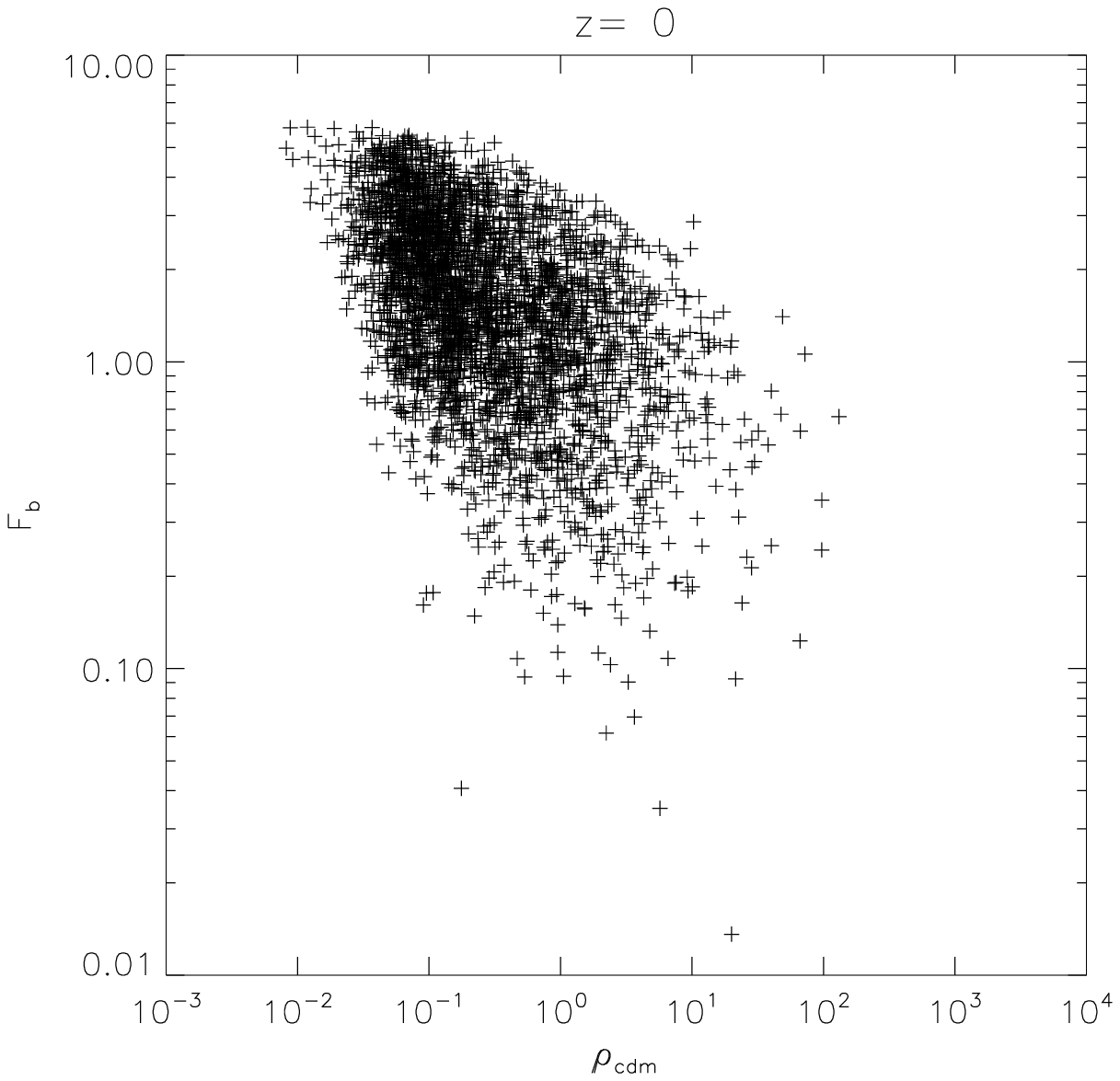}
\caption{$F_b$ vs. mass density of CDM, in unit of cosmic mean, of cells
randomly selected from
those in which $(1/2)\omega^2- S_{ij}S_{ij}$ is positive and $(1/2)\omega^2-
S_{ij}S_{ij}> |4\pi G(\rho_{tot}-\rho_0)|$.
 }
\end{figure*}

Figure 6 plots the $F_b$ vs. mass density of dark matter in randomly
selected cells in which $(1/2)\omega^2- S_{ij}S_{ij}$ is positive
and $(1/2)\omega^2- S_{ij}S_{ij}$ dominant over $|4\pi G(\rho_{\rm
tot}-\rho_0)|$. It shows clearly that the higher the density
$\rho_{\rm cdm}$ the lower the baryon fraction $F_b$. Since high
density $\rho_{\rm cdm}$ means high $|4\pi G(\rho_{\rm
tot}-\rho_0)|$, Figure 6 indicates that as long as the turbulence of
the IGM is dominant, the baryon fraction is lower for higher
$\rho_{\rm cdm}$.

We can also see from Figure 6 that a significant fraction of those
data point which have $F_b \leq 0.3$, i.e. more than $70$ percent of
baryonic matter are missing, locate in the areas with mass density
$\rho_{\rm cmd} \sim 1$. The event of $F_b \leq 0.3$ baryon
missing can occur even in the regions out of collapsed clumps,
indicating that the baryon missing of virialized halos should be
considered as result of the evolution of baryon distribution on
sizes larger than virialized halos.

\section[]{Baryon Fraction of dark matter halos}

\subsection{Turbulence pressure and halo mass}

It has been shown that the relation $P_{\omega}(k)=k^2P_{v}({\bf r})$
is well satisfied by the IGM velocity field in the scale range from
0.2 to 3 $h^{-1}$ Mpc at $z=0$ (Zhu et al 2010), on which the IGM is
in the state of fully developed homogeneous and isotropic turbulence.
The power spectrum of kinetic energy density of
the turbulence is approximately $E(k)\propto k^{-1.4}$. The
turbulence pressure (or the kinetic energy density) is weakly
scale-dependent $p_{\rm tur}(R)\propto R^{-0.4}$ and about the
order of magnitude $\sim 1\times 10^{-16}$ g cm$^{-1}$ s$^{-2}$.
On the other hand, the viral temperature $T_{200}$ of a halo at the
virial radius $r_{200}$ is $ \propto r_{200}^2$. The kinetic energy
density of the virial motion at $r_{200}$ is then $\sim 1 \times
10^{-16}(T_{200}/10^{5.5})$. The mass of halos with
$T_{200}=10^{5.5}$ K is about $10^{11}$ $h^{-1}$ $M_{\odot}$.
Obviously, the effect of turbulence pressure on the IGM collapsing
would be comparable with and even larger than the gravity for halos with
mass equal to and less than $10^{11}$ $h^{-1}$ $M_{\odot}$. The
baryon fraction would significantly be reduced for halos with mass
$\leq 10^{11}$ $h^{-1}$ $M_{\odot}$.

Meanwhile, the kinetic energy density of turbulence is small
than that of the virial motion of halos with mass larger than times
of $10^{11}$ $h^{-1}$ $M_{\odot}$. Thus, the halo mass dependence of
the baryon massing would show two phases: when the halo mass is
larger than times of $10^{11}$ $h^{-1}$ $M_{\odot}$, the
baryon fraction $f_b$ is not much affected by turbulent IGM, and it
will be substantially decreasing with halo mass when halo mass is
lower than a few $10^{11}$ $h^{-1}$ $M_{\odot}$.

\subsection{Halo mass dependence of baryon fraction}

With the preparation of above sections, we now analysis the
baryon fraction in gravitational collapsed halos. The accuracy of
the baryon fraction in selected halos relies on the calculation of
the mass of baryonic matter enclosed in the viral radius. To reduce
the system error, we divide each halo sphere into sub-cubic cells with
size equal to $1/50$ of the sphere radius. We then use
Cloud-In-Cell (CIC) interpolation to determine the baryon density at
each sub-cell. The halo mass $M_{200}$ and baryon mass are given
by the sum of the mass of total matter and baryonic matter respectively
in all the sub-cubic cells. This method
may yield large uncertainty for halos with $r_{200}$ less than the
size of a grid. To restrain the uncertainty, our statistical results mainly are on
$M_{200} \geq  10^{11} M_{\odot}$, i.e. the virial radius is about
equal to or larger than two grids.

Figure 7 displays the baryon fraction as a function of halo
mass at $z=0$, where the halos are identified by FOF (left) and HOP
(right) respectively. The results of the two panels are statistically the identical.
At a given halo mass, there are large scatters in the baryon fraction $F_b$, the
scatters are seen
to be  even more significant in the FOF algorithm than the HOP.
This is consistent with Figures 6, which indicates a large scatter in $F_b$
at a given mass density.  The distribution of $F_b$ is highly non-Gaussian and
has a long tail
on the side of small $F_b$, conforming with statistics of turbulent fluid, of
which the velocity and density fields are intermittent and their PDFs have long tails.

\begin{figure*}
\begin{center}
\includegraphics[width=7.0cm]{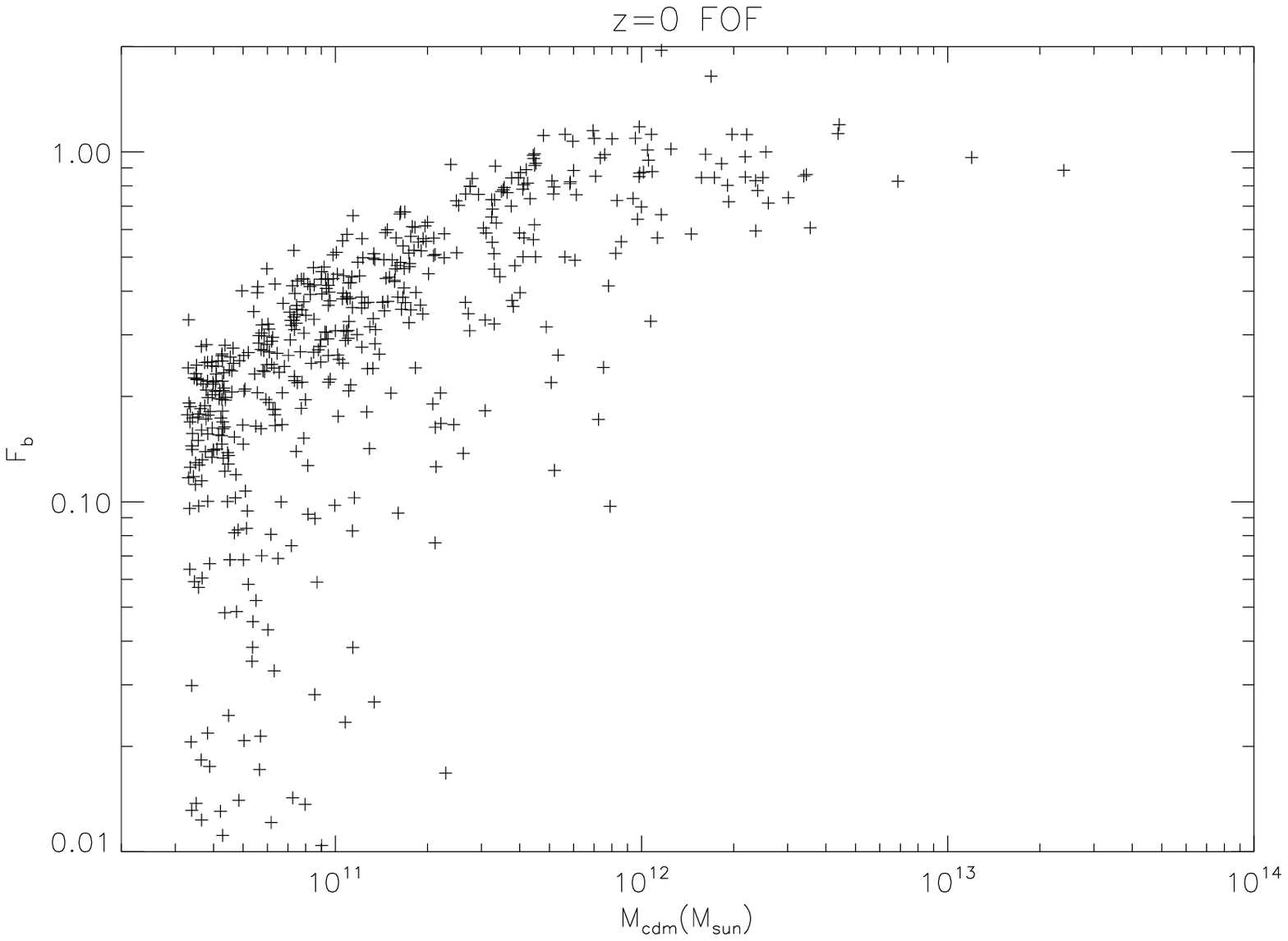}
\includegraphics[width=7.0cm]{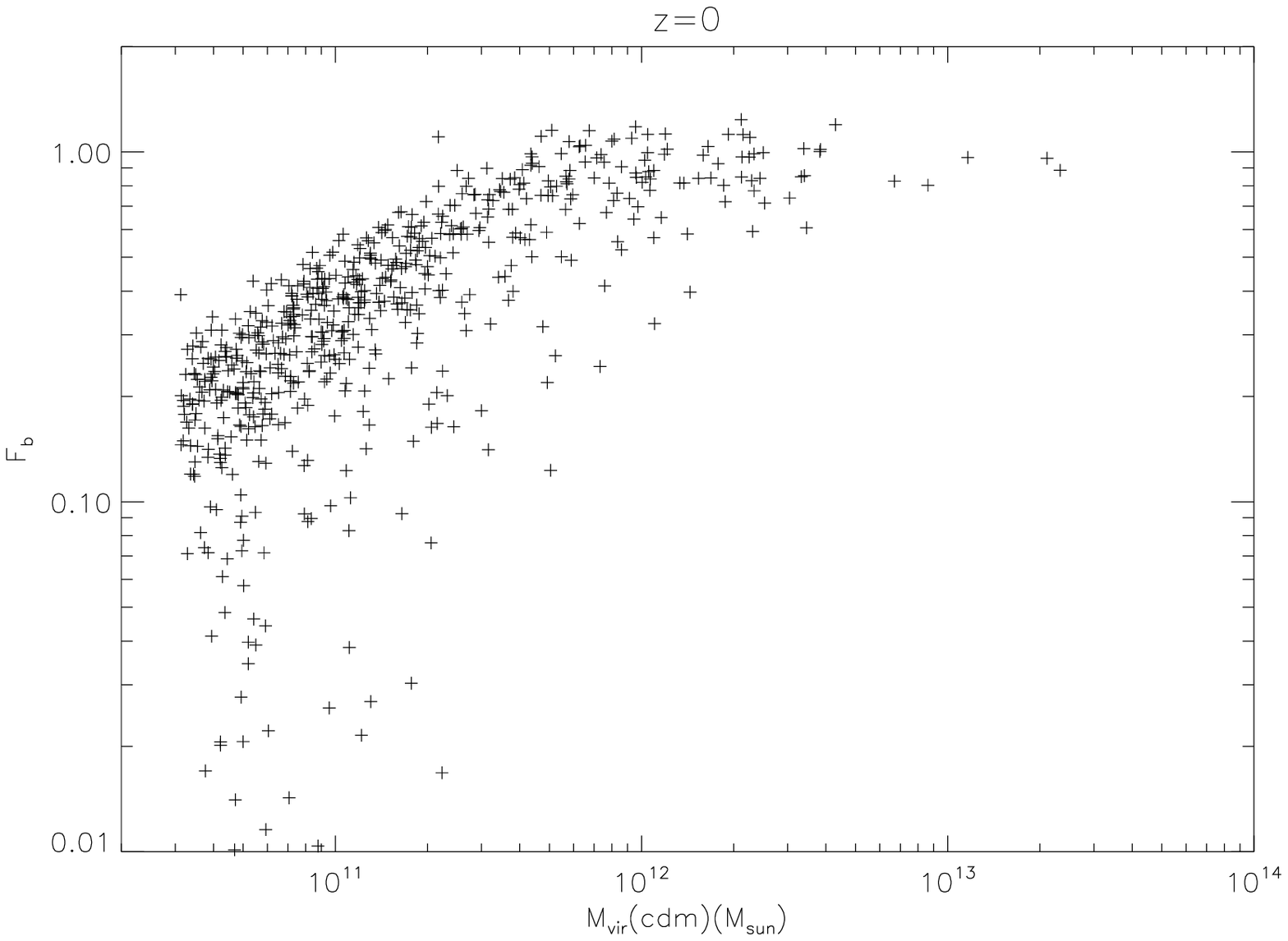}
\end{center}
\caption{Baryon fraction $F_b$ as function of halo mass $M_{200}$ at $z=0$. The halos
are identified with FOF method (left) and HOP method (right). }
\end{figure*}

Regardless of the scatters, the distribution of $F_b$ in Figure 7
can not be described by a single power law, but shows two distinct phases in
halo mass-dependence. In the mass range $10^{12}-10^{13}$ $h^{-1}$
$M_{\odot}$, $F_b$ is weakly dependent on halo mass and takes a value of
$80\%$ to $90\%$ of the cosmic mean, which is the first phase.
When halo mass is less than $\leq 10^{12}$ $M_{\odot}$, $F_b$ is quickly dropping off with
decreasing halo mass, i.e. the second phase. At mass$\sim 10^{11}$ $h^{-1}$ $M_{\odot}$,
$F_b$ is around $0.3$ of the cosmic mean. For halos with mass $M_{200}=3
\times 10^{10} M_{\odot}$, the mean of $F_b$ is even as low as about
$0.1$. Although the data points in the second phase shows a large
scattering, this phase has a clear upper envelop. The strong
downward  in  $F_b$ with decreasing halo mass is evident even taking account of the
scatters. The scatter looks very small on the side of high
halo mass.

\begin{figure*}
\begin{center}
\includegraphics[width=8cm]{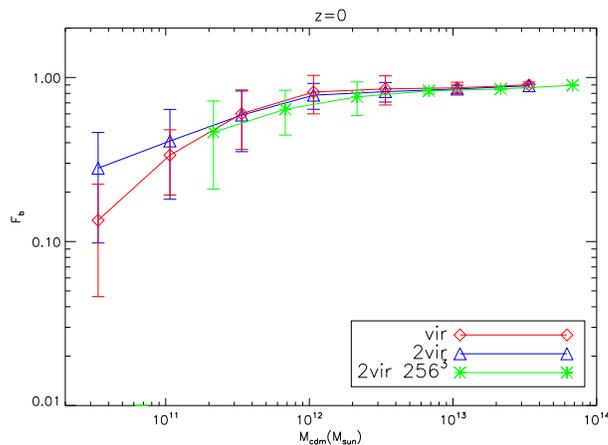}
\end{center}
\vspace{-5mm}
\caption{Baryon fraction $F_b$ as function of halo mass $M_{200}$. The $F_b$ is calculated
within the radius $r_{200}$ (red), 2$r_{200}$ (blue) for $512^{3}$ samples
and within 2$r_{200}$ for the 256$^3$ samples, respectively. The error bars
are the $rms$ of the scattered distribution.}
\end{figure*}

To test the convergence and stability of the results of Fig. 7, we
 1.) calculate $F_b$ within radius
2$r_{200}$; 2.) analysis 256$^3$ samples (\S 3). For the 256$^{3}$
samples, we only calculate the $F_{b}$ within radius 2$r_{200}$ to
avoid large system error, as $r_{200}$ is  under the grid size
at the low mass end. The results are given in Fig. 8, in which the
halos  are identified by the HOP algorithm. It shows that
$F_b$ in 2$r_{200}$ is about the same as that of $r_{200}$ within
the error bars for halos larger than times of $
10^{11}$ $h^{-1}$ $M_{\odot}$. For halos with mass $\leq
10^{11}$ $h^{-1}$ $M_{\odot}$, $F_b$ in $r_{200}$ shows a little
lower than that of 2$r_{200}$. This result seems to be reasonable, if
noted that the baryon fraction of cluster increases with radii (e.g.
Ettori \& Fabian 1999, Wu \& Xue 2000).

For the $256^{3}$ simulation, $F_{b}$ in radius 2$r_{200}$ shows
about $10 \%$ lower than its counterpart of the $512^{3}$ simulation
at halo mass $10^{12}-10^{11}$ $h^{-1}$ $M_{\odot}$. This difference
probably comes from the uncertainty caused by the size of grids. The uncertainty of $F_b$
caused by the resolution of $512^{3}$ simulation should be around $10 \%$ at
halo mass $10^{11}$ $h^{-1}$ $M_{\odot}$

Figure 9. compares the baryon fraction as a function of halo mass
$M_{200}$ at redshifts $z=1$, and 2. The halos in Figure 9 are
identified by HOP method, which yields similar
results as the FOF method. Clearly, it  shows the scatter in $F_b$ becomes
more significant at lower redshifts, data points have
$F_b<0.2$ at $z=2$ are quite few, but significantly
increases at $z=1$.

\begin{figure*}
\begin{center}
\includegraphics[width=6.5cm]{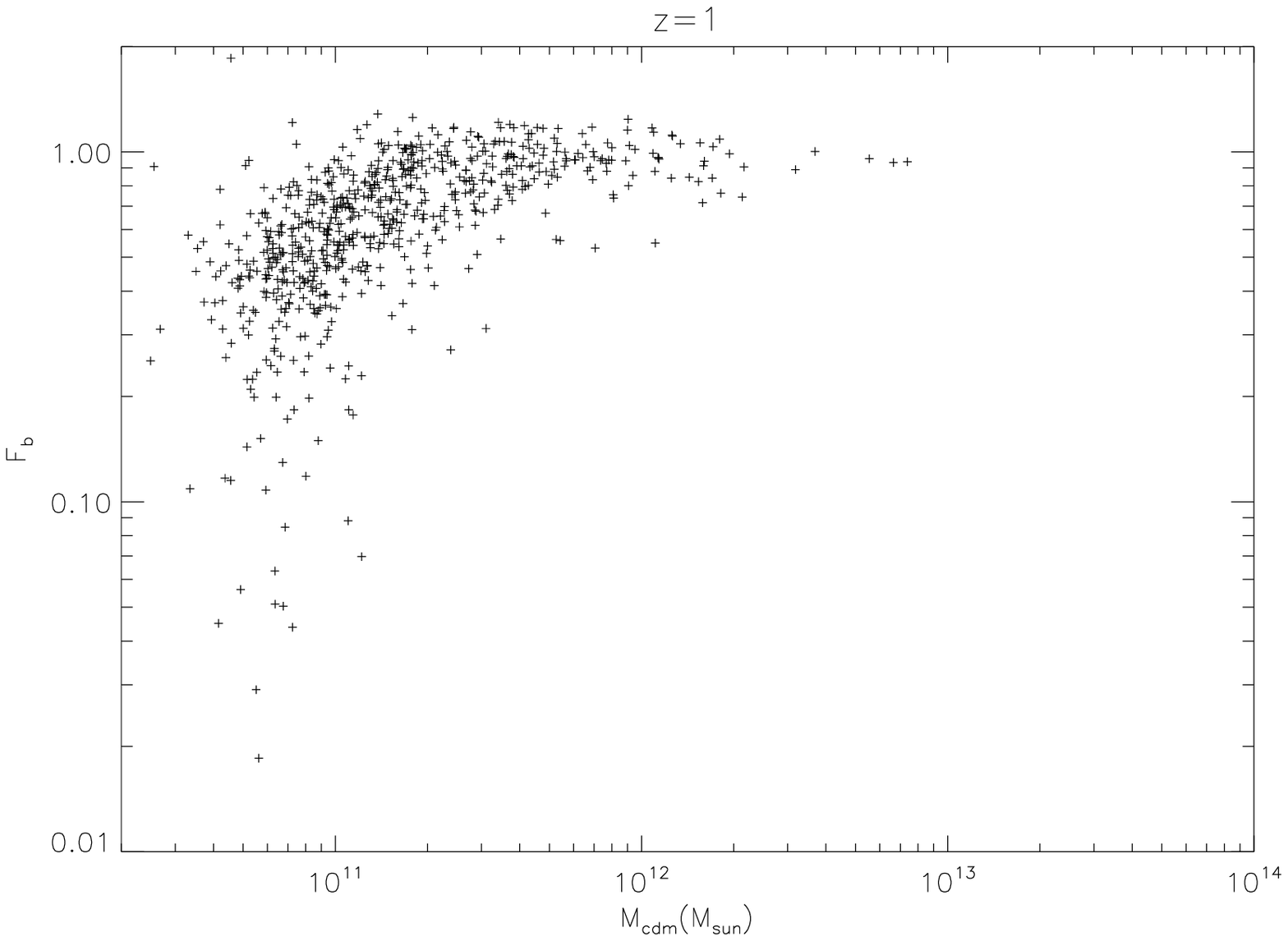}
\includegraphics[width=6.5cm]{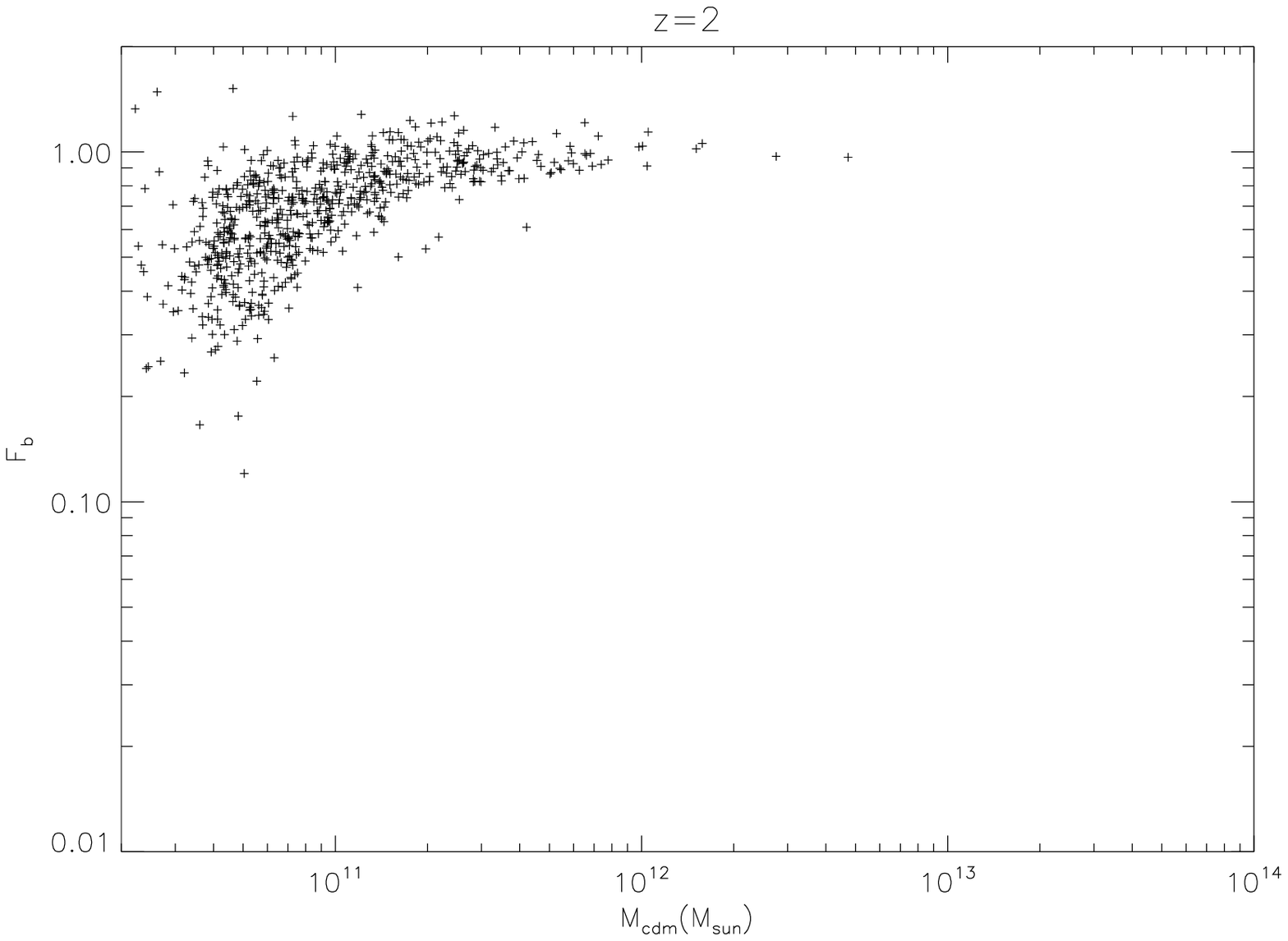}
\end{center}
\caption{Baryon fraction at $2r_{200}$ vs. halo mass $M_{200}$ at $z=1$(left) and
$z=2$(right). The halos are identified with HOP method.}
\end{figure*}

The redshift evolution of $F_b$ is parallel to
the IGM turbulence (Zhu et al 2010). The
turbulence is fully developed on scales from 0.2 $h^{-1}$ Mpc up to
a couple of Mpc since redshift $z\sim 2$. In the IGM density distribution
shown in Figure 1, there are about $7.6\%$ of volume with $(1/2)\omega^2-
S_{ij}S_{ij}>0$ at redshift $z=0$, while it is merely
$2.6\%$ at  redshift $z=2$. The effect of turbulence becomes
stronger to slow down the IGM clustering at lower redshifts.

\subsection{Comparison with observation }

Many measurements on the baryon fraction of galaxies, groups and
clusters have been done in recent years. Although it is widely
accepted that the baryon missing occurs in gravitational bound objects, the
result has not well quantitatively settled yet. For instance, some
measurements claim that the baryon fraction in most massive, relaxed
galaxy clusters is close to the cosmic mean (e.g. Vikhlinin et al. 2006;
Allen et al. 2008), while someone else argues that the baryon
fraction of clusters with mass $\sim 10^{14}$ M$_{\odot}$ is less
than the cosmic mean (Giodini et al. 2009). Nevertheless, observation
shows that the baryon fraction of objects with mass $\leq 10^{13}$
M$_{\odot}$ is systematically decreasing with decreasing masses.
For galaxy groups and galaxies, the baryon fraction is probably
not higher than about 0.1 of the cosmic mean (e.g. Sun et al. 2009;
Hoekstra et al. 2005; Heymans et al. 2006; Mandelbaum et al. 2006;
Gavazzi et al. 2007).

\begin{figure*}
\begin{center}
\includegraphics[width=8.0cm]{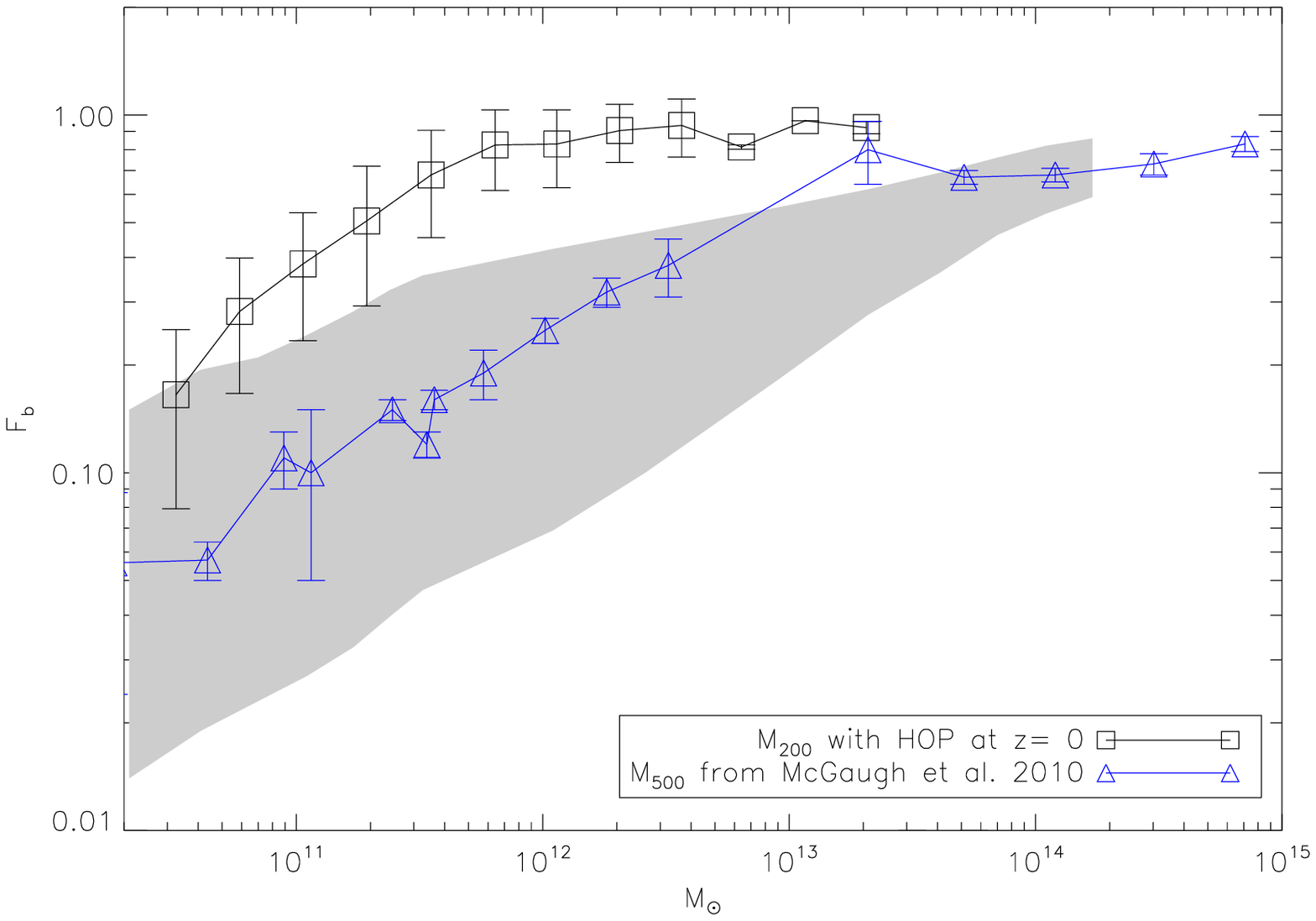}
\end{center}
\caption{Baryon fraction vs. mass of objects given by the data of McGaugh
et al. (2010) (blue); Dai et al (2010) (grey region), and simulation result
of $f_b$ at $r_{200}$ of the HOP halos at z=0 with a bin of 0.25 dex in mass.}
\end{figure*}

Figure 10 shows the  mass dependence of baryon fraction $F_b$
measured at $r_{500}$ (MaGaugh et al. 2010) and $r_{200}$ (Dai
et al (2010). These two data show some difference at mass $\geq
10^{13}$ $h^{-1}$ $M_{\odot}$. It is probably due to the cool gas is
underestimated with X-ray measurement. It may be also partially caused
by the difference in redshifts of the two sample sets.

The result of  MaGaugh et al.(2010) can not be fitted by a single
power law, but show two-phase feature. In the mass range $> 10^{13}$
$h^{-1}$ $M_{\odot}$,  $F_b\sim 0.85$, almost
independent of object mass. In the mass range $< 10^{13}$ $h^{-1}$
$M_{\odot}$, the baryon fraction decreases significantly with the
halo mass. $F_b$ is only about 10\% of $f_b^{\rm cosmic}$ at $
10^{11}$ $h^{-1}$ $M_{\odot}$ and further downs to 6\%
at halo mass of $3\times 10^{10}$ $h^{-1}$ $M_{\odot}$.

In Figure 10, we also plot the mean and variance of simulation
baryon fraction $F_b$ in $r_{200}$ for HOP halos at z=0. The
simulated mass-dependence of $F_b$ shows the similar trend as
observation. In the mass range of $ \sim 2\times 10^{11}- 10^{13}$
h$^{-1}$ M$_{\odot}$, simulation results are higher than observed
data by a factor of about 2, but within the error bars of observed
data. The effect of IGM turbulence is not the only reason
leading to the baryon missing. Yet it should be an important
dynamical reason of the baryon missing, especially for halos with
mass $\leq 2 \times 10^{11}$ $h^{-1}$ M$_{\odot}$.

\section[]{Discussion}

\subsection{Comparison with previously numerical studies}

The baryon depletion  has attracted many studies with
cosmological hydrodynamical simulation. As revealed in this paper,
the missing baryon could be caused by the IGM turbulence.  In order
to accurately study the effect of the turbulent IGM on baryon fraction
distribution and hence the baryon mission
, the algorithm of hydrodynamical simulation should be 
effective to capture curved shocks, vortices, and 
intermittence of turbulence in the IGM.

It has been shown that the smoothed particle hydrodynamics (SPH)
method may not be able to handle shocks or discontinuities as well
as grid method, because the nature of SPH is to smooth between 
particles(Tasker et al. 2008). The SPH method is found to have strong
damping of velocity fluctuations and fluid shear instabilities with
respect to grid method (Agertz et al. 2007). Moreover, the
artificial viscous force (dissipation) in SPH algorithm will 
reduce the Reynolds number, and suppress the effect of turbulence
(Dolag et al. 2005). These factors would be part of the reasons that
some SPH simulations find the baryon fraction in halos to be generally only $~10\%$ 
lower than the cosmic mean(e.g. Eke, Navarro \& Frenk 1998; Frenk et al. 1999;
Ettori 2006; Crain et al. 2007), where the lower in $f_{b}$ results from
energy transfer during shock formation(Navarro \& White 1993).

The Eulerian grid method, including both fixed grid and those with adaptive mesh
refinement(AMR), are shown to be effective to pick up the turbulent behavior of baryon
fluid in and around galaxies and clusters (e.g. Norman \&
Bryan 1999; Ryu et al 2008; Molnar et al 2009; Vazza et al. 2009;
Burns, Skillman\& O'Shea 2010;  Ruszkowsky \& Oh 2010). 
However, very few Eulerian simulation 
has been conducted on the turbulent behavior of IGM other than Zhu et al.(2010). 
O’Shea et al. (2005) investigated the baryon
fraction in halos at z = 3 during their study on the performance
comparison of ENZO and SPH method without looking into tur-
bulence. However, their small box, merely 3 Mpc, and relative
poorly resolved dark matter halos, typically tens to hundreds parti-
cle in a halo, makes a meaningful comparison between their results
and those presented in this paper unavailable. The energy cascade
would be suppressed in small box, which would significantly re-
strain the development of turbulence in the IGM. Very recently, the
turbulence of IGM is studied with grid based AMR code ENZO in-
cluding a subgrid scale model for small scale unresolved turbulence
in Iapichino et al. (2011), which yields many results similar to Zhu
et al (2010). However, statistical study on the baryon fraction is not
carried out.

The WENO algorithm embeded in our simulation uses a convex combination of all the
stencils in the reconstruction procedure, where each stencil is assigned with a
nonlinear weight depending on the local smoothness. This algorithm provides a high order accurate
way to capture the non-oscillatory property near strong
discontinuities as well as complex smooth solution features. This feature 
makes the WENO scheme has gained rapid popularity in the simulation of
complex-structure of hydrodynamical field (Shu 1999, 2009). The 5-th order 
WENO scheme used in this work makes it possible to follow high Reynolds
number baryonic fluid. On the other hand, like other high order non-artificial-viscosity
scheme, e.g. piecewise perturbation method
(PPM), with fixed grid, the computation time needed is much more than the
SPH and AMR method, which would become  more obvious in higher resolution work. 

\subsection{Systematic effects}

To estimate the effect of turbulence of IGM, a key factor is the
upper scale of fully developed turbulence. In some algorithm, the
energy of the turbulence is calculated with the fluctuations of
velocity with respect to the mean velocity in  cells with selected
size. An underestimated cell size generally leads to undervalued
turbulence (Dolag et al. 2005). In our algorithm the upper scale of
turbulence  is determined by the relation $P_{\rm
vor}(k)=k^2P_{v}(k)$, by which the ambiguity in selecting the cell
size is avoided.

Star formation and its feedback on the IGM evolution are not
considered in our simulation. The injection of hot gas and energy by
supernova explosions, AGN or other sources of cosmic rays is not
followed. These factors can be properly added, if the star
formation history and radiative transfer is well known, which in
turn need very high resolution. As mentioned above, the influence of
injecting hot gas and energy by supernovae is believed to yield
further decrease of $F_b$ in less massive halos, while feedback from
AGN might works in massive central galaxies. Including all these
mechanisms would give better consistency with observation.
Meanwhile, an accurate and robust observation result of baryon
fraction in objects from as massive as galaxy cluster  to low
mass dwarf galaxy is full of challenge. Large scatter in the
observed quantities and significant system bias comes from the
models applied in object structure and radiative transfer urges much
more efforts to deal with.

The spatial resolution of our simulation is $48.8 h^{-1}$kpc, which
is not enough to simulate virialized halos with mass less than
times of $2 \times 10^{10}$ $h^{-1}$ M$_{\odot}$. The lower limit of
the inertial range of turbulence may be affected by the grid resolution,
which is already seen in the vorticity power spectra given by
samples with different resolution in Zhu et al. (2010). On the other
hand, the box size of our simulation is $25 h^{-1}$Mpc. It may lose
the effect of long wavelength perturbations and hence underestimate
the nonlinear evolution of IGM velocity field. Nevertheless, we
believe that the basic dynamical picture features revealed by the
current samples would be valid when these factors are improved.

\section[]{Conclusion}

Since gravity is of scale free, it is generally believed that in the
scenario of hierarchical clustering, the formation and evolution of
halos is scale free in a large range of halo mass. However, the
dynamics of the system consisting of two components, dark matter and
IGM, is very different from those of one component system. In the
nonlinear regime, the hydrodynamical nature of the IGM leads to the
dynamical and statistical departure of the IGM from the dark matter.
The two component system is no longer to follow the scaling of
gravitational clustering of pure dark matter. This deviation is a
reason of the baryon missing in gravitational collapsed halos.

The dynamical equation of vorticity does not contain terms of
gravity. Therefore, the IGM turbulence characterized by the
vorticity will yield, at least, two scales, which violate the scale
free of the gravitational hierarchical clustering: 1.) the length
scale on which the IGM fluid has been developed to the state of
fully developed turbulence; 2.) the mass scale on which the
turbulence pressure is comparable with gravity of halo considered.
These scales play important role in the evolution of baryon
fraction. With cosmological hydrodynamic simulation, we find that at $z=0$ the first scale
is about 3 h$^{-1}$ Mpc, and the second one is $\sim 10^{11}$
h$^{-1}$ M$_{\odot}$. With these results,
we reach to the following conclusions:

i. The distribution of baryon fraction is highly nonuniform on
scales from hundreds kpc to a few of Mpc, and $f_b$ varies from as
low as 1\% to a few times of the cosmic mean.

i. The turbulence can effectively prevent the IGM from falling into
potential wells of dark matter halos with mass $\sim 10^{11}$
h$^{-1}$ M$_{\odot}$.

iii. The $f_b$ in dark matter halos is decreasing from $0.8f_b^{\rm
cosmic}$ at halo mass scales around $10^{12}$ $h^{-1} $ M$_{\odot}$
to $0.3f_b^{\rm cosmic}$ at $10^{11}$ h$^{-1}$ M$_{\odot}$ due to
the turbulent state of the IGM.

The estimated turbulence pressure at $z=0$ correspond to a random
motion with r.m.s velocity of about $50-100$ km $s^{-1}$ in the
scale range from hundreds of kpc and up to $\sim 2$ Mpc. The
turbulent pressure is dynamical and non-thermal. When the turbulence
dissipated, its kinetic energy becomes the thermal energy. It yields
the entropy in halos (He, Feng, \& Fang 2005). Therefore, the
dissipation of turbulence actually is a mechanism of heating, which
gives a compensation to the cooling of gas in halos. This result is
consistent with the Burgers' shock heating (He et al 2004).
In summmary, the dynamics of turbulence can effectively affect the
baryon fraction of halos .
\section[]{Acknowledgments}

WSZ acknowledges the support of the International Center for
Relativistic Center Network (ICRAnet). Our work takes advantage of
open facilities provided by the NASA HPCC ESS group at the
University of Washington (http://www-hpcc.astro.washington.edu/).
This work is partially supported by the National Science Foundation
of China grant NSFC 10633040,  10725314 , 10621303 and the  973
Program under contract No. 2007CB815402.

\appendix

\section{Shock capturing algorithm of WENO scheme}

In the WENO scheme, the smoothness of a hydrodynamical quantities $U=(\rho,p,v,E)$
in any dimension is measured by
\begin{equation}
\eta_{i} = \frac{1}{4}(U_{i+1}-U_{i-1})^{2}+\frac{13}{12}(U_{i+1}-2U_{i}+U_{i-1})^{2},
\end{equation}
where the definition of $U_{i}$ is the same as that given in the Appendix of Zhu et al (2010).

With $\eta_{i}$, we can construct the one-dimensional discontinuous detector by
\begin{equation}
DDF_{i}=\left |\frac{(\eta_{i})^{1/2}}{U_{i+1}+2U_{i}+U_{i-1}}\right |^{n}.
\end{equation}
where $n=1,2,..$. We use $U= p$ and $n=2$ to detect the discontinuous surface (Visbal
\& Gaitonde 2005; Lo et al. 2007). For each cell, we calculate $DDF_i$ in three directions.
The value of $DDF$ is normalized by the max of $DDF$ among all the cells. Then, a
threshold parameter of $1.0^{-4}$ of $DDF_{i}$ is used to flag discontinuous
zone with $DDF_{i}>10^{-4}$. Finally, we pick up the shocks with Mach number $M_a>1.5$
from the flagged cells by  checking the conditions as that given by Ryu et al. 2003.

\end{document}